\documentclass[aps]{revtex4}
\usepackage{amsfonts}
\usepackage{amsmath}
\usepackage{amssymb,epsf}

\begin{document}

\title{Extended phase space thermodynamics and $P-V$ criticality:\\
Brans-Dicke-Born-Infeld vs Einstein-Born-Infeld-dilaton black
holes}
\author{S. H. Hendi$^{1,2}$\footnote{email address:
hendi@shirazu.ac.ir}, R. Moradi Tad$^{1}$, Z. Armanfard$^{1}$ and
M. S. Talezadeh$^{1}$} \affiliation{$^1$ Physics Department and
Biruni Observatory, College of Sciences, Shiraz
University, Shiraz 71454, Iran\\
$^{2}$ Research Institute for Astronomy and Astrophysics of
Maragha (RIAAM), Maragha, Iran}

\begin{abstract}
Motivated by thermodynamic analogy of black holes and Van der
Waals liquid/gas system, in this paper, we study $P-V$ criticality
of both dilatonic Born-Infeld black holes and their conformal
solutions, Brans-Dicke-Born-Infeld solutions. Due to the conformal
constraint, we have to neglect the old Lagrangian of dilatonic
Born-Infeld theory and its black hole solutions, and introduce a
new one.  We obtain spherically symmetric nonlinearly charged
black hole solutions in both Einstein and Jordan frames and then,
we calculate the related conserved and thermodynamic quantities.
After that, we extend the phase space by considering the
proportionality of the cosmological constant and thermodynamical
pressure. We obtain critical values of thermodynamic coordinates
through numerical methods and plot relevant $P-V$ and $G-T$
diagrams. Investigation of the mentioned diagrams helps us to
study thermodynamical phase transition. We also analyze the
effects of varying different parameters on the phase transition of
black holes.
\end{abstract}

\maketitle

\section{Introduction\label{Intro}}

In $1872$ James Clerk Maxwell combined the electricity and
magnetism laws in a unified theory. In Maxwell's formulation of
electromagnetism, the electric field of a point-like charge is
singular at its position which leads to infinite self-energy. In
$1934$ Max Born and Leopold Infeld \cite{m1.3569,2.1412.3569}
introduced a new theory with an upper bound of the electric field
in order to obtain finite value for the self-energy of point-like
charges in the classical electrodynamics.

In theoretical physics, the Born--Infeld (BI) model is known as an
example of an interesting and valuable nonlinear electrodynamics
\cite{m1.3569,2.1412.3569}. BI action has obtained vast attentions
for many reasons. For example, in the context of superstring
theory, the low energy dynamics of D-branes is handled by BI
action \cite{m3.3569}. In addition, loop correction analysis of
quantum field theory leads to BI type action
\cite{m4.3569,m5.3569,m6.3569}. BI electrodynamics also exhibits
some physical properties regarding wave propagation like the
absence of shock waves \cite{wave1,wave2}. This theory is very
close to Einstein's idea of introducing a nonsymmetrical metric
with an antisymmetric part as the electromagnetic field and a
symmetric part as the usual metric. It can also be regarded as a
covariant generalization of Mie's theory \cite{groupBI}.
Einstein-BI theory has led to some interesting observable
predictions in the context of solar interior dynamics, big bang
nucleosynthesis, neutron stars \cite{m23,25}, the nonsingular
cosmological models and alternative to inflation \cite{m26}. The
supersymmetric version of the BI Lagrangian is constructed in
Refs. \cite{m6,m7}, while in Refs. \cite{m8,m9,m10} it was
identified as an invariant action of the Goldstone multiplet in
$N=2$ supersymmetric theory which is spontaneously broken to
$N=1$. The results of Ref. \cite{m9} has been generalized to the
case of $n$ vector multiplets in $N=2$ supersymmetry
\cite{m11,m12} with explicit solution for $n=2$ and $n=3$.
Moreover, the scalar perturbation at the pre-inflationary stage
driven by a massive scalar field in Eddington-inspire BI gravity
was investigated in Ref. \cite{m13}.

One of the main reason for considering scalar-tensor theory is
that it may give a clue to interpret the acceleration expansion of
the Universe \cite{Acceleration}. In addition, inflation may be
explained based on the scalar-tensor theory of gravitation. Such
inflationary model is known as the hyper-extended inflation
\cite{Inflation}. On the cosmological point of view, inflation can
be naturally accommodated in the (low-energy limit of) string
theory, since it contains fundamental scalar field which acts as
inflation. In addition, at high energy regimes it seems that
gravity cannot describe by the Einstein theory, but may be
modified by the superstring terms. Such modified superstring terms
contain dilaton field.

On the other hand, the theory of Brans-Dicke (BD) is one of the
modified theories of general relativity that gets convenient data
of several cosmological problems like inflation, early and late
behavior of the universe, coincidence problem and cosmic
acceleration \cite{m9m,m2.07858b}. It was shown that BD theory can
be used for dark energy modeling \cite{m9m,m10m}. This theory has
been probed to be a possible explanation for the accelerated
expansion of our universe \cite{m6.08315a,m6.08315b,m6.08315c}.

The BD theory is a modified form of general relativity that is
made by coupling a scalar field with a gravitational tensor field.
It has a free constant, known as the coupling parameter $\omega$
(tuneable parameter), which can be adjusted according to the
suitable observational evidences. The four dimensional action of
BD theory has the following form \cite{m1504.06723}
\begin{equation*}
S=\frac{1}{16\pi }\int d^{4}x\sqrt{-g}[\phi R-\frac{\omega }{\phi
}\partial _{\mu }\phi \partial ^{\mu }\phi ].
\end{equation*}

This theory is compatible with weak equivalence principle, Mach's
principle and Dirac's large number hypothesis \cite{m3.07858}. In
addition, it is in agreement with solar system experimental
observations for specific domain of $\omega$ \cite{m4.07858}. BD
theory has been considered in various branches of gravitation and
cosmology. Instability analysis of the Schwarzschild black hole in
BD gravity is discussed in Ref. \cite{m9.07858}. The analytical
and numerical features of static spherically symmetric solutions
in the context of BD-like cosmological model have been explored
\cite{m1504.06723}. Thermodynamical properties of higher
dimensional charged rotating black brane solutions in BD-BI
gravity are presented in Ref. \cite{m.0808}.The cosmological
perturbation equations of BD-BI have been investigated in Ref.
\cite{Amendola}. The interaction between two test objects and the
anomalous acceleration of Pioneer 10 and Pioneer 11 spacecraft
\cite{Pioneer10,Pioneer11} have been studied in BD-BI context
\cite{Smolyakov}. Higher dimensional BI-dilaton black hole
solutions with nonabelian Yang-Mills field and its stability
against linear radial perturbations have been examined in
\cite{BIYMd}.

There are several reasons for considering the cases in which the
scalar field is non-minimally coupled to the BI field. Here we
present some of them. Non-minimal coupling between a scalar fields
and a nonlinear $U(1)$ gauge field emerges naturally in
supergravity and in the low-energy effective action of string
theory. Following the works of Refs. \cite{m4.3569,BId2,BId3}, it
was shown that the BI action coupled to a dilaton field appears in
the low energy limit of open superstring theory. In other words,
considering the coupling of gravity to other gauge fields, the
presence of the dilaton field cannot be ignored. Therefore on the
electrodynamic point of view, one remains with Einstein–BI-dilaton
gravity or its conformal transformation, BD-BI theory. In
addition, asymptotically adS (nonlinearly) charged dilaton black
hole solutions may known as the family of four-charge black holes
in $N = 8$ four dimensional gauged supergravity
\cite{supergravity}. Furthermore, thermodynamical phase structure
of dilatonic (BI) black objects is more interesting with respect
to Reissner--Nordstr\"{o}m solutions \cite
{Cadoni2010,Charmousis2010,Doneva2010,m.0808}. Also, in the
holographic perspective, (nonlinearly) charged dilatonic black
holes have shown a rather rich and interesting phenomenology \cite
{Cadoni2010,Gubser2010,Goldstein2010A,Goldstein2010B,Chen2010,Lee2010B,Liu2010,Lee2010C}.
Moreover, regarding gauge/gravity duality, nonlinearly charged
black holes with a non-minimally coupled to a scalar field are
good candidates for gravitational side dual to Lifshitz-like
theories \cite{Goldstein2010A,Bertoldi2010,Perlmutter2011}.

Regarding black hole thermodynamics as an important connection
between quantum gravity and classical nature of general relativity
\cite{m1505.05517}, one may be motivated to consider thermal phase
transition. The behavior of a thermodynamic system can be
explained with physical temperature and entropy
\cite{m1.03340shA,m1.03340shB}. On the other hand, by considering
the extended phase space of black hole, we may treat the
cosmological constant proportional to a typical dynamical pressure
\cite{m3.0352shA,m3.0352shB,m3.0352shC,m3.0352shD,m3.0352shE,m3.0352shF,m3.0352shG,m3.0352shH,m3.0352shI,m3.0352shJ,m3.0352shK,m3.0352shL,m3.0352shM,m3.0352shN}.
Considering thermodynamics of black holes, we should note that
phase transition plays an important role in describing critical
phenomena from the thermodynamics and quantum points of view. The
thermodynamic behavior of charged black hole solutions in BD
theory and the analogy of these solutions with the Van der Waals
liquid-gas system in the extended phase space was investigated in
Ref. \cite{Armanfard}. In this paper, we study the $P-V$
criticality and phase transition of charged black holes in BD-BI
theory and compare it with Einstein-BI-dilaton solutions. At first
we consider the Lagrangian of Einstein-BI-dilaton gravity and
investigate its thermodynamical properties. Then, we present a
brief discussion regarding the conformal inconsistency of this
Lagrangian and introduce a new well-defined Lagrangian of
Einstein-BI-dilaton theory. We obtain its exact solutions and
analyze thermodynamic behavior of black holes. We also use the
conformal transformation to obtain correct BD-BI black hole
solutions. Finally, we discuss $P-V$ criticality of the black
holes in both Einstein and Jordan frames.

\section{Part A: OLD LAGRANGIAN:}

\subsection{Black hole solutions in Einstein-BI-dilaton gravity}

The well-known action of $(n+1)$- dimensional BI-dilaton gravity
is
\begin{equation}
I=\int_{\mathcal{M}}d^{n+1}x\sqrt{-g}\left(
\mathcal{R}-\frac{4}{n-1}(\nabla \Phi )^{2}-V(\Phi)+L\left(
\mathcal{F},\Phi \right) \right) ,  \label{I1}
\end{equation}%
where $\mathcal{R}$ is the Ricci scalar, $V(\Phi )$ is a
self-interacting potential for scalar field $\Phi $ and
$L(\mathcal{F},\Phi )$ is the coupled Lagrangian of BI-dilaton
theory \cite{sheykhi-riazi}
\begin{equation}
L\left(\mathcal{F},\Phi \right) =4\beta ^{2}e^{4\alpha {\Phi
}/\left( n-1\right) }\left( 1-\sqrt{1+\frac{e^{-8\alpha {\Phi
}/\left( n-1\right) }\mathcal{F}}{2\beta ^{2}}}\right) .
\label{LFPhi}
\end{equation}%

In Eq. (\ref{LFPhi}), the constants $\alpha$ and $\beta$ are,
respectively, dilaton and BI parameters, $\mathcal{F}=F_{{\mu \nu
}}F^{\mu \nu }$ is the Maxwell invariant, in which $F_{\mu \nu
}=\partial _{\lbrack \mu }A_{\nu ]}$ and $A_{\mu }$ is the gauge
potential. In the limit of $\beta \rightarrow \infty $, $L\left(
\mathcal{F},\Phi \right)$ reduces to the standard Maxwell
Lagrangian coupled to a dilaton field
\begin{equation*}
L(\mathcal{F},\Phi) =-e^{-4\alpha {\Phi }/\left( n-1\right)}
\mathcal{F}.
\end{equation*}

Variation of the action (\ref{I1}) with respect to $g_{\mu \nu }$,
$\Phi $ and $F_{{\mu \nu }}$ leads to the following field
equations \cite{sheykhi-riazi}
\begin{eqnarray}
R_{\mu \nu } &=&\frac{4}{n-1}\left( \partial _{\mu }\Phi \partial _{\nu
}\Phi +\frac{1}{4}g_{\mu \nu }V(\Phi )\right) -4e^{-4\alpha \Phi
/(n-1)}\partial _{Y}L\left( Y\right) F_{\mu \eta }F_{\nu }^{\eta }+  \notag
\\
&&\frac{4\beta ^{2}}{\left( n-1\right) }e^{4\alpha \Phi /\left( n-1\right)
}[2Y\partial _{Y}L\left( Y\right) -L\left( Y\right) ]g_{\mu \nu },
\label{FE1Sh}
\end{eqnarray}%
\begin{equation}
\nabla ^{2}\Phi =\frac{n-1}{8}\frac{\partial V}{\partial \Phi }+2\beta
^{2}\alpha e^{-4\alpha \Phi /(n-1)}\left[ 2Y\partial _{Y}L\left( Y\right)
-L\left( Y\right) \right] ,  \label{FE2Sh}
\end{equation}%
\begin{equation}
\partial _{\mu }\left[ \sqrt{-g}e^{-4\alpha \Phi /(n-1)}\partial _{Y}L\left(
Y\right) F^{\mu \nu }\right] =0,  \label{FE3Sh}
\end{equation}%
where we have used the following notations
\begin{equation*}
L(\mathcal{F},\Phi )=4\beta ^{2}e^{4\alpha \Phi /(n-1)}L\left(
Y\right) ,
\end{equation*}%
\begin{equation*}
L(Y)=1-\sqrt{1+Y},
\end{equation*}%
\begin{equation*}
Y=\frac{e^{\frac{-8\alpha \Phi }{n-1}}\mathcal{F}}{2\beta ^{2}}.
\end{equation*}

In order to obtain static solutions, one can assume the following metric
\begin{equation}
ds^{2}=-f(r)dt^{2}+\frac{dr^{2}}{f(r)}+r^{2}R(r)^{2}d\Omega
_{k}^{2},
\end{equation}%
where%
\begin{equation}
d\Omega _{k}^{2}=\left\{
\begin{array}{cc}
d\theta _{1}^{2}+\sum\limits_{i=2}^{n-1}\prod\limits_{j=1}^{i-1}\sin
^{2}\theta _{j}d\theta _{i}^{2} & k=1 \\
d\theta _{1}^{2}+\sinh ^{2}\theta _{1}d\theta _{2}^{2}+\sinh ^{2}\theta
_{1}\sum\limits_{i=3}^{n-1}\prod\limits_{j=2}^{i-1}\sin ^{2}\theta
_{j}d\theta _{i}^{2} & k=-1 \\
\sum\limits_{i=1}^{n-1}d\phi _{i}^{2} & k=0%
\end{array}%
\right. ,  \label{dOmega}
\end{equation}%
indicates the Euclidean metric of an $(n-1)$-dimensional
hypersurface with constant curvature $(n-1)(n-2)k$ and volume
$\varpi _{n-1}$. It has been shown that the following functions
satisfy all field equations simultaneously \cite{sheykhi-riazi}
\begin{eqnarray}
f(r) &=&-\frac{k(n-2)(1+\alpha ^{2})^{2}}{(\alpha ^{2}-1)(\alpha ^{2}+n-2)}%
\left( \frac{r}{b}\right) ^{2\gamma }-\frac{m}{r^{(n-1)(1-\gamma )-1}}+\frac{%
2\Lambda (1+\alpha ^{2})^{2}r^{2}}{(n-1)(\alpha ^{2}-n)}\left( \frac{r}{b}%
\right) ^{-2\gamma }-  \notag \\
&&\frac{4\beta ^{2}(\alpha ^{2}+1)^{2}r^{2}}{(n-1)(\alpha ^{2}-n)}\left(
\frac{r}{b}\right) ^{-2\gamma }\left\{ 1-\text{ }_{2}F_{1}\left( \left[ -%
\frac{1}{2},\frac{\alpha ^{2}-n}{2(n-1)}\right] ,\left[ \frac{\alpha ^{2}+n-2%
}{2(n-1)}\right] ,-\Delta \right) \right\} ,  \label{f(r)sh}
\end{eqnarray}%
\begin{equation}
R(r)=e^{\frac{2\alpha \Phi }{n-1}},  \label{Rsh}
\end{equation}%
\begin{equation}
\Phi(r)=-\frac{(n-1)\alpha }{2(1+\alpha ^{2})}\ln \left( \frac{r}{b}\right),
\label{Phish}
\end{equation}%
where $\gamma =\alpha ^{2}/(1+\alpha ^{2})$, $\Delta =\frac{q^{2}}{\beta
^{2}r^{2(n-1)}}\left( \frac{r}{b}\right) ^{2\gamma (n-1)}$ , $m$ is an
integration constant which is related to the total mass and $b$ is another
arbitrary constant related to the scalar field. In addition, one should
consider the following suitable Liouville-type potential \cite{sheykhi-riazi}%
\begin{equation}
\mathbf{V}(\Phi )=2\Lambda \exp \left( \frac{4\alpha \Phi }{n-1}\right) +%
\frac{k(n-1)(n-2)\alpha ^{2}}{b^{2}\left( \alpha ^{2}-1\right) }\exp \left(
\frac{4\Phi }{(n-1)\alpha }\right) ,  \label{VdSh}
\end{equation}%
to obtain consistent solutions. Calculation of curvature scalars
shows that there is a curvature divergency at the origin and all
curvature invariants are finite for $r\neq 0$
\cite{sheykhi-riazi}. It means that there is a physical
singularity located at $r=0$ which can be covered with an event
horizon, and therefore, one can interpret it as a black hole. In
the next section we study thermodynamics and $P-V$ criticality of
the mentioned solutions.

\subsubsection{\textbf{Extended phase space and P-V criticality of the
solutions}}

Thermodynamic properties of the mentioned solutions were discussed
before \cite{sheykhi-riazi}. In this section, we study the $P-V$
criticality in the extended phase space. At first, we calculate
the Hawking temperature by
using the surface gravity interpretation ($\kappa $)%
\begin{equation*}
T=\frac{\kappa }{2\pi }=\frac{1}{2\pi }\sqrt{-\frac{1}{2}\left( \nabla _{\mu
}\chi _{\nu }\right) \left( \nabla ^{\mu }\chi ^{\nu }\right) }=\frac{%
f^{\prime }(r_{+})}{4\pi },
\end{equation*}%
where $\chi =\partial /\partial t$ is the null Killing vector of the event
horizon $r_{+}$.\ Calculations lead to the following explicit relation \cite%
{sheykhi-riazi}
\begin{eqnarray}
T &=&\frac{-\left( n-2\right) \left( 1+\alpha ^{2}\right) }{2\pi (\alpha
^{2}+n-2)r_{+}}\left( \frac{r_{+}}{b}\right) ^{2\gamma -1}+\frac{(n-\alpha
^{2})m}{4\pi (1+\alpha ^{2})}r_{+}^{(n-1)(\gamma -1)}+\frac{q^{2}\left(
1+\alpha ^{2}\right) r_{+}^{3-2n}}{\pi \left( \alpha ^{2}+n-2\right) }\left(
\frac{r_{+}}{b}\right) ^{2\gamma \left( n-2\right) }\times  \notag \\
&&_{2}F_{1}\left( \left[ \frac{1}{2},\frac{\alpha ^{2}+n-2}{2\left(
n-1\right) }\right] ,\left[ \frac{\alpha ^{2}+3n-4}{2\left( n-1\right) }%
\right] ,-\Delta _{+}\right) ,  \label{Tsh}
\end{eqnarray}%
where $\Delta _{+}=\frac{q^{2}\left( \frac{r_{+}}{b}\right) ^{2\gamma \left(
n-1\right) }}{\beta ^{2}r_{+}^{2(n-1)}}$. The entropy and the finite mass of
the black hole can be obtained with the following forms \cite{sheykhi-riazi}
\begin{equation}
S=\frac{\varpi _{n-1}b^{(n-1)\gamma }}{4}r_{+}^{(n-1)\left(
1-\gamma \right)},  \label{Ssh}
\end{equation}%
\begin{equation}
M=\frac{\varpi _{n-1}b^{(n-1)\gamma }}{16\pi }\left( \frac{n-1}{1+\alpha ^{2}%
}\right) m,  \label{Msh}
\end{equation}%
where $m$ is the (geometrical) mass parameter of the black hole
which can be expressed in term of the event horizon radius
\begin{eqnarray}
\left. m\right\vert _{r=r_{+}} &=&-\frac{\left( n-2\right) \left( \alpha
^{2}+1\right) ^{2}r_{+}^{-\gamma \left( n-1\right) +n-2}}{\left( n+\alpha
^{2}-1\right) \left( \alpha ^{2}-1\right) }\left( \frac{r_{+}}{b}\right)
^{2\gamma }+\frac{2\Lambda \left( 1+\alpha ^{2}\right) ^{2}b^{2\gamma
}r_{+}^{-\gamma \left( n+1\right) +n}}{\left( n-1\right) \left( \alpha
^{2}-n\right) }-  \notag \\
&&\frac{4\beta ^{2}\left( 1+\alpha ^{2}\right) ^{2}b^{2\gamma
}r_{+}^{-\gamma \left( n+1\right) +n}}{\left( n-1\right) \left( \alpha
^{2}-n\right) }\left\{ 1-\text{ }_{2}F_{1}\left( \left[ -\frac{1}{2},\frac{%
\alpha ^{2}-n}{2\left( n-1\right) }\right] ,\left[ \frac{\alpha ^{2}+n-2}{%
2\left( n-1\right) }\right] ,-\Delta _{+}\right) \right\} .  \label{mh}
\end{eqnarray}

Now, we extend the phase space by defining a thermodynamical
pressure proportional to cosmological constant and its
corresponding conjugate quantity as the volume. Using the concept
of energy-momentum tensor, one can find the generalized definition
for the pressure in the presence of dilaton field \cite{Armanfard}
\begin{equation}
P=-\frac{\Lambda }{8\pi }\left( \frac{b}{r_{+}}\right) ^{2\gamma
}, \label{Psh}
\end{equation}%
where in the absence of dilaton field ($\alpha =\gamma =0$), the known
relation $P=\frac{-\Lambda }{8\pi }$ is recovered.

By regarding the relation between cosmological constant and
thermodynamical pressure, one may interpret the mass as enthalpy.
Hence we can calculate the generalized volume as
\begin{equation}
V=\left( \frac{\partial H}{\partial P}\right) _{S,Q}=\left( \frac{\partial M%
}{\partial P}\right) _{S,Q}=\frac{\varpi _{n-1}\left( 1+\alpha
^{2}\right) r_{+}^{n}}{n-\alpha ^{2}}\left( \frac{b}{r_{+}}\right)
^{\gamma \left( n-1\right) },  \label{Vsh}
\end{equation}%
where in the absence of dilaton field, one obtains $V=\frac{\varpi
_{n-1}r_{+}^{n}}{n}$, as expected.

Now, we are in a position to study the phase transition through
$P-V$ and $G-T$ diagrams. The equation of state of the black hole
can be written, using Eqs. (\ref{Tsh}) and (\ref{Psh}) in the
following form
\begin{equation}
P=\frac{\left( n-2\right) \left( n-1\right) }{16\pi \left( \alpha
^{2}-1\right) }\left( \frac{r_{+}}{b}\right) ^{2\gamma }r_{+}^{-2}+\frac{%
\left( n-1\right) T}{4\left( 1+\alpha ^{2}\right) r_{+}}+\frac{\beta ^{2}}{%
4\pi }\left( \frac{b}{r_{+}}\right) ^{2\gamma }\left( \sqrt{1+\Delta _{+}}%
-1\right) .  \label{Psh2}
\end{equation}

Due to the relation between the volume and radius of the black
hole, we use the horizon radius (specific volume) in order to
investigate the critical behavior of these systems
\cite{m3.0352shH,m3.0352shI,m3.0352shJ}. Considering the mentioned
equation of state, one can investigate the inflection point of
$P-r_{+}$ diagram to obtain the phase transition point. The
inflection point of isothermal curves in $P-r_{+}$ diagram has the
following properties
\begin{equation}
\left( \frac{\partial P}{\partial r_{+}}\right) _{T}=0,
\label{dPdV}
\end{equation}%
\begin{equation}
\left( \frac{\partial ^{2}P}{\partial r_{+}^{2}}\right) _{T}=0.
\label{d2PdV2}
\end{equation}

One can use Eqs. (\ref{dPdV}) and (\ref{d2PdV2}) with the equation
of state (\ref{Psh2}) to calculate the critical values for
temperature, pressure and volume. In addition, we can study the
phase transition through calculation of Gibbs free energy. Since
we are working in the extended first law of black hole
thermodynamics, $M$ (finite mass of the black hole) will be
interpreted as enthalpy instead of internal energy, and therefore,
the Gibbs free energy of the black hole can be written as
\begin{equation}
G=H-TS=M-TS,  \label{G1}
\end{equation}%
where after some manipulations, we obtain the following Gibbs free
energy per unit volume $\varpi _{n-1}$
\begin{eqnarray}
G &=&\frac{\left( \alpha ^{4}-1\right) \beta ^{2}r_{+}^{n}\left( 1-\sqrt{%
1+\Delta _{+}}\right) }{4\pi \left( n-1\right) \left( n-\alpha
^{2}\right) \left( \frac{r_{+}}{b}\right) ^{\gamma \left(
n+1\right) }}+\frac{\left( 1+\alpha ^{2}\right) \left( n-2\right)
r_{+}^{n-2}}{16\pi \left( n+\alpha ^{2}-2\right) }\left(
\frac{b}{r_{+}}\right)^{\gamma \left( n-3\right) }
\notag \\
&&+\frac{\left( \alpha ^{4}-1\right) P r_{+}^{n}}{\left(
n-1\right) \left( n-\alpha ^{2}\right) }\left(
\frac{b}{r_{+}}\right) ^{\gamma \left( n-1\right) }+\frac{\left(
1+\alpha ^{2}\right) \left( n-1\right) q^{2}\sqrt{1+\Delta
_{+}}}{4\pi \left( n-\alpha ^{2}\right) \left( n+\alpha
^{2}-2\right) r_{+}^{n-2} }\left( \frac{r_{+}}{b}\right) ^{\gamma
(n-3)}\times  \notag \\
&&_{2}F_{1}\left( \left[ 1,1+\frac{\alpha ^{2}-1}{2\left(
n-1\right) }\right] ,\left[ \frac{\alpha ^{2}+3n-4}{2\left(
n-1\right) }\right] ,-\Delta _{+}\right) .  \label{Gsh}
\end{eqnarray}%

We should note that the characteristic swallow-tail behavior in
$G-T$ diagrams guarantees the existence of the phase transition.

\subsection{Black hole solutions in BD-BI gravity}

In this section, we discuss the possibility of BD-BI solutions
which are conformally related to the obtained BI-dilaton black
holes. To do so, one should find a suitable conformal
transformation to obtain BD-BI counterpart of action (\ref{I1}).
In the action of BD theory, dilaton field should be decoupled from
matter field (electrodynamics) and be coupled with gravity. In
other words, one should find a suitable conformal transformation
in which it transforms action (\ref{I1}) to the following known
BD-BI action \cite{sh5}
\begin{equation}
I_{BD-BI}=-\frac{1}{16\pi }\int_{\mathcal{M}}d^{n+1}x\sqrt{-g}\left( \Phi
\mathcal{R}\text{ }-\frac{\omega }{\Phi }(\nabla \Phi )^{2}-V(\Phi )+%
\mathcal{L}(\mathcal{F})\right) ,  \label{acBD}
\end{equation}%
where $\mathcal{L}(\mathcal{F})$ is the Lagrangian of BI theory \cite{m.0808}
\begin{equation}
\mathcal{L}(\mathcal{F})=4\beta ^{2}\left( 1-\sqrt{1+\frac{\mathcal{F}}{%
2\beta ^{2}}}\right) .  \label{abBD}
\end{equation}

In Eq. (\ref{acBD}) $\mathcal{R}$ is the Ricci scalar, $\omega $
is the coupling constant, $\Phi $ denotes the BD scalar field and
$V(\Phi)$ is a self--interaction potential for $\Phi$. It is
notable that in the limit of $\beta \rightarrow \infty $,
$\mathcal{L}(\mathcal{F})$ reduces to the standard Maxwell form
$\mathcal{L}(\mathcal{F})=-\mathcal{F}$, as it should.

It is a matter of calculation to show that there is no consistent
conformal transformation between actions (\ref{I1}) and
(\ref{acBD}) \cite{HendiTalezadeh}. In other words, although the
action (\ref{I1}) leads to the known action of Maxwell-dilaton
gravity for $\beta \rightarrow \infty $, it is not conformally
related to the BD-BI action, and therefore, Eq. (\ref{I1}) is not
a suitable and consistent action. Considering the conformally
ill-defined action (\ref{I1}), in the next section, we follow the
method of Ref. \cite{HendiTalezadeh} to obtain a conformally
well-defined action of BI-dilaton gravity.

\section{Part B: New LAGRANGIAN: \newline FIELD EQUATIONS AND CONFORMAL TRANSFORMATIONS \label{FE}}

The action of $(n+1)$- dimensional BD-BI theory with a scalar
field $\Phi $ and a self-interacting potential $V(\Phi )$ can be
written as Eq. (\ref{acBD}). Variation of this action with respect
to $g_{\mu \nu }$, $\Phi $ and $F_{{\mu \nu }}$ leads to
\cite{sh5}
\begin{eqnarray}
G_{\mu \nu } &=&\frac{\omega }{\Phi ^{2}}\left( \nabla _{\mu }\Phi
\nabla
_{\nu }\Phi -\frac{1}{2}g_{\mu \nu }(\nabla \Phi )^{2}\right) -\frac{V(\Phi )%
}{2\Phi }g_{\mu \nu }+\frac{1}{\Phi }\left( \nabla _{\mu }\nabla
_{\nu }\Phi
-g_{\mu \nu }\nabla ^{2}\Phi \right)   \notag \\
&&+\frac{2}{\Phi }\left( \frac{F_{\mu \lambda }F_{\nu }^{\text{ }\lambda }}{%
\sqrt{1+\frac{\mathcal{F}}{2\beta ^{2}}}}+\frac{1}{4}g_{\mu \nu }\mathcal{L}(%
\mathcal{F})\right) ,  \label{FBD1} \\
\nabla ^{2}\Phi  &=&\frac{1}{2\left[ \left( n-1\right) \omega +n\right] }%
\left( (n-1)\Phi \frac{dV(\Phi )}{d\Phi }-\left( n+1\right) V(\Phi
)+\left(
n+1\right) \mathcal{L}(\mathcal{F})+\frac{4 \mathcal{F}}{\sqrt{1+\frac{\mathcal{F}}{%
2\beta ^{2}}}}\right) ,  \label{FBD2}
\end{eqnarray}

\begin{equation}
\nabla _{\mu }\left( \frac{F^{\mu \nu
}}{\sqrt{1+\frac{\mathcal{F}}{2\beta ^{2}}}}\right) =0.
\label{FBD3}
\end{equation}

Due to the appearance of the scalar field in the denominator of
field equation (\ref{FBD1}), solving Eqs.
(\ref{FBD1})-(\ref{FBD3}), directly, is a non-trivial task. In
order to remove this difficulty, one can use the traditional
approach; the conformal transformation. Indeed, using the
conformal transformation \cite{sh5}, the BD-BI theory will be
transformed into the Einstein--BI--dilaton gravity. The suitable
conformal transformation is as follows
\begin{equation}
\bar{g}_{\mu \nu }=\Phi ^{2/(n-1)}g_{\mu \nu },  \label{CT}
\end{equation}%
where%
\begin{eqnarray}
\bar{\Phi} &=&\frac{n-3}{4\alpha }\ln \Phi ,  \label{Phibar} \\
\alpha  &=&(n-3)/\sqrt{4(n-1)\omega +4n}.  \label{alpha}
\end{eqnarray}

Applying the mentioned conformal transformation, one finds that
the action of BD-BI and its related field equations change to the
well-known dilaton gravity with the following explicit forms
\begin{equation}
\overline{I}_{G}=-\frac{1}{16\pi }\int_{\mathcal{M}}d^{n+1}x\sqrt{-\overline{g}}%
\left\{ \overline{\mathcal{R}}-\frac{4}{n-1}(\overline{\nabla}\overline{\Phi})^{2}-\overline{V}(%
\overline{\Phi})+\overline{L}\left(
\overline{\mathcal{F}},\overline{\Phi }\right) \right\} ,
\label{con-ac}
\end{equation}%
\begin{eqnarray}
\overline{\mathcal{R}}_{\mu \nu } &=&\frac{4}{n-1}\left( \overline{\nabla}_{\mu }\overline{%
\Phi}\overline{\nabla}_{\nu }\overline{\Phi}+\frac{1}{4}\overline{V}(\overline{\Phi})\overline{g}%
_{\mu \nu }\right) -\frac{1}{n-1}\overline{L}(\overline{\mathcal{F}},\overline{\Phi })%
\overline{g}_{\mu \nu }+\frac{2e^{-\frac{4\alpha \overline{\Phi }}{n-1}}}{%
\sqrt{1+\overline{Y}}}\left( \overline{F}_{\mu \eta
}\overline{F}_{\nu }^{\eta
}-\frac{\overline{\mathcal{F}}}{n-1}\overline{g}_{\mu \nu }\right)
,
\label{FE1} \\
\overline{\nabla}^{2}\overline{\Phi} &=&\frac{n-1}{8}\frac{\partial \overline{V}(\overline{\Phi})%
}{\partial \overline{\Phi}}+\frac{\alpha }{2(n-3)}\left( (n+1)\overline{L}(\overline{\mathcal{F}}%
,\overline{\Phi })+\frac{4e^{-\frac{4\alpha \overline{\Phi
}}{n-1}}\overline{\mathcal{F}}}{\sqrt{1+\overline{Y}}}\right) ,
\label{FE2}
\end{eqnarray}
\begin{equation}
\overline{\nabla }_{\mu }\left( \frac{e^{-\frac{4\alpha \overline{\Phi }}{n-1%
}}}{\sqrt{1+\overline{Y}}}\overline{F}^{\mu \nu }\right) =0,
\label{FE3}
\end{equation}%
where $\overline{\mathcal{R}}$ and $\overline{\nabla}$ are,
respectively, the Ricci scalar and covariant differentiation
related to the metric $\overline{g}_{\mu \nu }$. In addition, the
potential $\overline{V}\left( \overline{\Phi }\right) $ and
the BI-dilaton coupling Lagrangian $\overline{L}\left( \overline{F},%
\overline{\Phi }\right) $ are, respectively \cite{HendiTalezadeh}
\begin{equation}
\overline{V}(\overline{\Phi})=\Phi ^{-(n+1)/(n-1)}V(\Phi ),
\label{poten}
\end{equation}%
and
\begin{equation}
\overline{L}\left( \overline{\mathcal{F}},\overline{\Phi }\right)
=4\beta ^{2}e^{-4\alpha \left( n+1\right) \overline{\Phi }/\left[
\left( n-1\right) \left( n-3\right) \right] }\left(
1-\sqrt{1+\frac{e^{16\alpha \overline{\Phi
}/\left[ \left( n-1\right) \left( n-3\right) \right] }\overline{\mathcal{F}}}{%
2\beta ^{2}}}\right) .  \label{LFP}
\end{equation}

It is easy to find that $\overline{L}\left( \overline{\mathcal{F}},\overline{\Phi }%
\right) \rightarrow 0$ as $\beta \rightarrow 0$, and on the other
side, it reduces to the following standard Maxwell-dilaton
Lagrangian in the limit of $\beta \rightarrow \infty $
\begin{equation}
\overline{L}\left( \overline{\mathcal{F}},\overline{\Phi }\right)
=-e^{-4\alpha \overline{\Phi }/\left( n-1\right)
}\overline{\mathcal{F}}.  \label{LFPmax}
\end{equation}

In other words, comparing both Lagrangians of BI-dilaton theory
(Eqs. (\ref{LFPhi}) and (\ref{LFP})), one finds that although both
Lagrangians lead to Maxwell-dilaton Lagrangian for the weak
nonlinearity strength ($\beta \rightarrow \infty $), only the new
Lagrangian (Eq. (\ref{LFP})) is the correct and consistent one
with conformal transformation.

It is notable that we used the following notations for writing the
field equations (\ref{FE1})-(\ref{FE3}).
\begin{equation}
\overline{L}\left( \overline{\mathcal{F}},\overline{\Phi }\right)
=4\beta ^{2}e^{-4\alpha \left( n+1\right) \overline{\Phi }/\left[
\left( n-1\right) \left( n-3\right) \right]
}\overline{L}\overline{(Y}), \label{LFP2}
\end{equation}%
where
\begin{eqnarray}
\overline{L}\left( \overline{Y}\right) &=&1-\sqrt{1+\overline{Y}},
\label{L(Y)} \\
\overline{Y} &=&\frac{e^{16\alpha \overline{\Phi }/\left[ \left(
n-1\right) \left( n-3\right) \right]
}\overline{\mathcal{F}}}{2\beta ^{2}}. \label{Y}
\end{eqnarray}

Taking into account the conformal relation of BD-BI theory and
BI-dilaton gravity, one can find that if $\left( \overline{g}_{\mu
\nu },\overline{F} _{\mu \nu },\overline{\Phi }\right) $ is the
solution of Eqs. ( \ref{FE1})-(\ref{FE3}) with potential
$\overline{V}(\overline{\Phi}) $, then
\begin{equation}
\left[ g_{\mu \nu },F_{\mu \nu },\Phi \right] =\left[ \exp \left( -\frac{%
8\alpha \overline{\Phi}}{\left( n-1\right) (n-3)}\right)
\overline{g}_{\mu \nu },\overline{F}_{\mu \nu },\exp \left(
\frac{4\alpha \overline{\Phi}}{n-3}\right) \right]  \label{BDsol}
\end{equation}
is the solution of BD-BI field equations (Eqs.
(\ref{FBD1})-(\ref{FBD3})) with potential $V(\Phi)$.

\section{Part C: New LAGRANGIAN: EXACT SOLUTIONS}

\subsection{Black hole solutions in Einstein-BI-dilaton gravity and BD-BI
theory \label{Sol}}

\subsubsection{\textbf{Einstein frame:}}

In this section, first we obtain the solutions of dilaton gravity
in the Einstein frame and then we use the conformal transformation
to obtain the solutions of the BD-BI theory.

We assume the following metric
\begin{equation}
d\overline{s}^{2}=-Z(r)dt^{2}+\frac{dr^{2}}{Z(r)}+r^{2}R^{2}(r)d\Omega
_{k}^{2}, \label{metric}
\end{equation}%
where $d\Omega _{k}^{2}$\ was presented in Eq. (\ref{dOmega}). We
obtain the consistent dilaton field as well as metric functions.
To do this, we should consider a potential
$\mathbf{\overline{V}}(\overline{\Phi})$. It was shown that a
suitable potential is the Liouville-type potential with BI
correction
\cite{HendiTalezadeh}%
\begin{equation}
\mathbf{\overline{V}}(\overline{\Phi})=2\Lambda \exp \left( \frac{4\alpha \overline{\Phi}}{%
n-1}\right) +\frac{k(n-1)(n-2)\alpha ^{2}}{b^{2}\left( \alpha ^{2}-1\right) }%
\exp \left( \frac{4\overline{\Phi}}{(n-1)\alpha }\right)
+\frac{W(r)}{\beta ^{2}}, \label{liovilpoten}
\end{equation}%
which reduces to $2\Lambda $ in the absence of dilaton field. In
other words, the first two terms of Eq. (\ref{liovilpoten}) come
from Maxwell-dilaton gravity \cite{Armanfard}, while the third
term appears because of the nonlinearity of electrodynamics (BI
effect).
Now, considering the potential (\ref{liovilpoten}) with Eqs. (\ref{FE1})- (%
\ref{FE3}), one finds
\begin{eqnarray}
F_{tr} &=&E(r)=\frac{qe^{\left( \frac{4\alpha \overline{\Phi }(r)}{n-1}%
\right) }}{(rR(r))^{(n-1)}\sqrt{1+\frac{e^{(\frac{8\alpha \overline{\Phi }(r)%
}{n-3})}q^{2}(rR(r))^{-2(n-1)}}{\beta ^{2}}}},  \label{E} \\
\overline{\Phi} &=&\frac{(n-1)\alpha }{2(1+\alpha ^{2})}\ln \left(
\frac{b}{r}\right) \label{phi}
\end{eqnarray}%
\begin{equation}
W(r)=\frac{4q(n-1)\beta ^{2}R(r)}{\left( 1+\alpha ^{2}\right)
r^{\gamma
}b^{n\gamma }}\int \frac{E(r)}{r^{n(1-\gamma )-\gamma }}dr+\frac{4\beta ^{4}%
}{R(r)^{\frac{2(n+1)}{n-3}}}\left( 1-\frac{E(r)R(r)^{(n-3)}}{qr^{1-n}}%
\right) -\frac{4q\beta ^{2}E(r)}{r^{n-1}}(\frac{r}{b})^{\gamma
(n-1)}, \label{W}
\end{equation}%
\begin{eqnarray}
Z(r) &=&-\frac{k\left( n-2\right) \left( \alpha ^{2}+1\right)
^{2}c^{-2\gamma }r^{2\gamma }}{\left( \alpha ^{2}+n-2\right)
\left( \alpha
^{2}-1\right) }+\left( \frac{(1+\alpha ^{2})^{2}r^{2}}{(n-1)}\right) \frac{%
2\Lambda \left( \frac{r}{b}\right) ^{-2\gamma }}{(\alpha ^{2}-n)}-\frac{m}{%
r^{(n-1)(1-\gamma )-1}}  \notag \\
&&-\frac{4(1+\alpha ^{2})^{2}q^{2}(\frac{r}{b})^{2\gamma
(n-2)}}{(n-\alpha ^{2})r^{2(n-2)}}\left( \frac{1}{2(n-1)}\digamma
_{1}(\eta )-\frac{1}{\alpha ^{2}+n-2}\digamma _{2}(\eta )\right) ,
\label{f}
\end{eqnarray}%
where $m$ and $b$ are integration constants related to the mass
and scalar field, respectively, and
\begin{eqnarray*}
\digamma _{1}(\eta ) &=&\text{ }_{2}F_{1}\left( \left[ \frac{1}{2},\frac{%
(n-3)\Upsilon }{\alpha ^{2}+n-2}\right] ,\left[ 1+\frac{(n-3)\Upsilon }{%
\alpha ^{2}+n-2}\right] ,-\eta \right) , \\
\digamma _{2}(\eta ) &=&\text{ }_{2}F_{1}\left( \left[ \frac{1}{2},\frac{%
(n-3)\Upsilon }{2(n-1)}\right] ,\left[ 1+\frac{(n-3)\Upsilon
}{2(n-1)}\right]
,-\eta \right) , \\
\Upsilon  &=&\frac{\alpha ^{2}+n-2}{2\alpha ^{2}+n-3}, \\
\eta  &=&\frac{q^{2}(\frac{r}{b})^{2\gamma
(n-1)(n-5)/(n-3)}}{\beta
^{2}r^{2(n-1)}}, \\
R(r) &=&\exp \left( \frac{2\alpha \overline{\Phi }}{n-1}\right)
=\left( \frac{r}{b}\right) ^{-\gamma }.
\end{eqnarray*}

It is notable that for $\beta \rightarrow \infty $, the last term
of Eq. (\ref{liovilpoten}) vanishes and the resultant relations
reduce to dilatonic Maxwell solutions \cite{Armanfard}.
Calculations show that the curvature scalars diverge at the origin
and they are finite for $r\neq 0$. In other words, since the
metric function $Z(r)$ has real positive root(s), one can
interpret the singularity as a black hole.

In the next section, we use the conformal transformation to obtain
BD-BI black hole solutions.

\subsubsection{\textbf{Jordan frame:}}

Now, we are going to obtain charged black hole solutions of BD-BI
theory. Using the conformal transformation (\ref{poten}), the
potential $\mathbf{V}(\Phi )$ in Jordan frame is
\begin{equation}
\mathbf{V}(\Phi )=2\Lambda \Phi ^{2}+\frac{k(n-1)(n-2)\alpha ^{2}}{%
b^{2}\left( \alpha ^{2}-1\right) }\Phi ^{\lbrack (n+1)(1+\alpha
^{2})-4]/[(n-1)\alpha ^{2}]}+\Phi ^{(n+1)/(n-1)}\frac{W(r)}{\beta
^{2}}. \label{V(phi)}
\end{equation}

Taking into account the solutions in Einstein frame with the
mentioned conformal transformation, we can obtain the solutions of
Eqs. (\ref{FBD1})-(\ref{FBD3}). Considering the following
$(n+1)-$dimensional metric
\begin{equation}
ds^{2}=-A(r)dt^{2}+\frac{dr^{2}}{B(r)}+r^{2}H^{2}(r)d\Omega
_{k}^{2}, \label{metric1}
\end{equation}%
we find that the functions $A(r)$ and $B(r)$ are
\begin{eqnarray}
A(r) &=&\left( \frac{r}{b}\right) ^{4\gamma /\left( n-3\right)
}Z\left(
r\right) ,  \label{A(r)} \\
B(r) &=&\left( \frac{r}{b}\right) ^{-4\gamma /\left( n-3\right)
}Z\left(
r\right) ,  \label{B(r)} \\
H(r) &=&\left( \frac{r}{b}\right) ^{-\gamma (\frac{n-5}{n-3})},
\label{H(r)}
\\
\Phi \left( r\right)  &=&\left( \frac{r}{b}\right)
^{-\frac{2\gamma \left( n-1\right) }{n-3}}.  \label{Phi}
\end{eqnarray}

Since the curvature scalars of the mentioned metric diverge at
$r=0$, it is easy to show that the corresponding solution can be
interpreted as black hole and it can be covered by an event
horizon which is the largest real root of $Z(r)$. Since it was
shown that for $k=0,-1$ there is no phase transition
\cite{Armanfard}, hereafter, we choose the positive curvature
constant boundary ($k=1$) to investigate phase transition.

\subsection{Thermodynamic properties and $P-V$ criticality: \\
dilatonic-BI vs BD-BI black holes \label{p-vb}}

\subsubsection{\textbf{Thermodynamic properties:}}

In this section, we focus on the conserved and thermodynamic
quantities of the black hole solutions in both Einstein and Jordan
frames. In order to calculate the Hawking temperature, one can use
the surface gravity
interpretation%
\begin{equation}
T=\frac{\kappa }{2\pi }=\left\{
\begin{array}{cc}
\frac{Z^{\prime }(r_{+})}{4\pi }, & \text{dilatonic BI} \\
\frac{1}{4\pi }\sqrt{\frac{B(r)}{A(r)}}A^{\prime }(r_{+}), & \text{BD-BI}%
\end{array}%
\right. .  \label{Td1}
\end{equation}

It is easy to show that Hawking temperature in Einstein frame is
exactly equal to that in Jordan frame with the following form
\begin{equation}
T=\frac{\left( \alpha ^{2}+1\right) }{2\pi \left( n-1\right) }\left[ -\frac{%
\left( n-2\right) (n-1)}{2\left( \alpha ^{2}-1\right) r_{+}}\left( \frac{%
r_{+}}{b}\right) ^{2\gamma }-\Lambda r_{+}\left(
\frac{r_{+}}{b}\right) ^{-2\gamma }+\Gamma _{+}\right] ,\text{
dilatonic BI \& BD-BI}  \label{Td2}
\end{equation}%
where
\begin{eqnarray}
\Gamma _{+} &=&-\frac{\left( \alpha ^{2}+1\right) ^{2}q^{2}}{2\pi (n-1)}%
\left( \frac{r_{+}}{b}\right) ^{2\gamma \left( n-2\right)
}r_{+}^{3-2n}\digamma _{1}(\eta _{+}),  \label{GAMMA} \\
\eta _{+} &=&\eta (r=r_{+}).
\end{eqnarray}

Since the conformal transformation is a regular smooth function at
the horizon, this equality is expected. In addition, the finite
mass and the entropy of the black hole for both frames can be
obtained with the following forms
\begin{eqnarray}
M &=&\frac{\varpi _{n-1}b^{(n-1)\gamma }}{16\pi }\left(
\frac{n-1}{1+\alpha
^{2}}\right) m,  \label{Md} \\
S &=&\frac{\varpi _{n-1}b^{(n-1)\gamma }}{4}r_{+}^{(n-1)\left(
1-\gamma \right) }.  \label{Sd}
\end{eqnarray}

Using the Gauss's law, the electric charge would have the following form%
\begin{equation}
Q=\frac{q}{4\pi },  \label{Qd}
\end{equation}%
which is valid for both frames.

Now, we extend the phase space by defining a thermodynamical
pressure proportional to the cosmological constant and its
corresponding conjugate quantity as the volume. Following the
method of \cite{Armanfard}, one finds the generalized definition
for the pressure in the presence of dilaton field as
\begin{equation}
P=-\frac{\Lambda }{8\pi }\times \left\{
\begin{array}{cc}
\left( \frac{r_{+}}{b}\right) ^{-2\gamma }, & \text{dilatonic BI} \\
\left( \frac{r_{+}}{b}\right) ^{-\frac{2\gamma \left( n-1\right)
}{n-3}}, &
\text{BD-BI}%
\end{array}%
\right. .  \label{Pd}
\end{equation}

We should note that in the absence of dilaton field ($\alpha
=\gamma =0$), the known relation $P=\frac{-\Lambda }{8\pi }$ is
recovered. In order to calculate the volume, we should obtain the
enthalpy. Following the previous interpretation of mass and
enthalpy, we can calculate the generalized volume as
\begin{equation}
V=\frac{\varpi _{n-1}\left( 1+\alpha ^{2}\right) r_{+}^{n}}{n-\alpha ^{2}}%
\left\{
\begin{array}{cc}
\left( \frac{r_{+}}{b}\right) ^{-\gamma \left( n-1\right) }, & \text{%
dilatonic BI} \\
\left( \frac{r_{+}}{b}\right) ^{-\frac{\gamma (n^{2}-4n-1)}{n-3}}, & \text{%
BD-BI}%
\end{array}%
\right. ,  \label{Vd}
\end{equation}%
where for $\alpha \rightarrow 0$, we obtain $V=\frac{\varpi _{n-1}r_{+}^{n}}{%
n}$, as expected. It is worthwhile to mention that although $P$
and $V$ are different for the Einstein and Jordan frames, Both
frames have the same multiplication of "$P \times V$".

\subsubsection{\textbf{$P-V$ criticality of dilatonic BI vs BD-BI:}}

Now, we are in a position to study the phase transition through
$P-V$ and $G-T$ diagrams. The equation of state of the black hole
can be written, using Eqs. (\ref{Td1}), (\ref{Pd}) and (\ref{Vd})
with the following form
\begin{equation}
P=\left[ \frac{(n-1)(n-2)}{16\pi (\alpha ^{2}-1)r_{+}^{2}}(\frac{r_{+}}{b}%
)^{2\gamma }+\frac{(n-1)T}{4(1+\alpha
^{2})r_{+}}+\frac{q^{2}}{8\pi
r_{+}^{2(n-1)}}(\frac{r_{+}}{b})^{2\gamma (n-2)}\digamma _{1}(\eta _{+})%
\right] \Xi ,  \label{P}
\end{equation}%
where
\begin{equation*}
\Xi =\left\{
\begin{array}{cc}
1, & \text{dilatonic BI} \\
(\frac{r_{+}}{b})^{-\frac{4\gamma }{n-3}}, & \text{BD-BI}%
\end{array}%
\right. .
\end{equation*}

Now, one can consider the mentioned equation of state to obtain
critical quantities through the properties of the inflection point
of $P-r_{+}$ diagram (Eqs. (\ref{dPdV}) and (\ref{d2PdV2})). In
addition, we can use Eq. (\ref{G1}) to obtain the Gibbs free
energy of black holes. After some manipulations, we obtain the
Gibbs free energy per unit volume $\varpi _{n-1}$ as
\begin{eqnarray}
G &=&\frac{(1+\alpha ^{2})(n-2)}{16\pi (\alpha ^{2}+n-2)}(\frac{r_{+}}{b}%
)^{-\gamma (n-3)}r_{+}^{n-2}+\frac{(\alpha ^{4}-1)q^{2}}{8\pi
(\alpha ^{2}-n)(n-1)}(\frac{r_{+}}{b})^{\gamma
(n-3)}r_{+}^{(2-n)}\digamma _{1}(\eta
_{+})  \notag \\
&&-\frac{(n-1)(1+\alpha ^{2})q^{2}}{4\pi (\alpha ^{2}-n)(\alpha ^{2}+n-2)}(%
\frac{r_{+}}{b})^{\gamma (n-3)}r_{+}^{(2-n)}\digamma _{2}(\eta _{+})+\frac{%
P(1-\alpha ^{4})r_{+}^{n}}{(\alpha ^{2}-n)(n-1)}\mathcal{G},
\label{G2}
\end{eqnarray}%
where
\begin{equation*}
\mathcal{G}=\left\{
\begin{array}{cc}
\begin{array}{c}
(\frac{r_{+}}{b})^{-\gamma (n-1)}, \\
\\
\end{array}
&
\begin{array}{c}
\text{dilatonic BI} \\
\\
\end{array}
\\
\begin{array}{c}
(\frac{r_{+}}{b})^{\frac{2\gamma (n-1)}{(n-3)}-\gamma (n+1)}, \\
\end{array}
& \text{BD-BI}%
\end{array}%
\right. .
\end{equation*}

As we mentioned before, existence of the characteristic
swallow-tail behavior in $G-T$ diagrams helps us to obtain thermal
phase transition (for further details, see  $G-T$ diagrams and
their related texts). Since analytical investigation of the
critical behavior is not possible, we use numerical analysis
instead. We present various tables and figures to study such
behavior.


\begin{figure}[tbp]
$%
\begin{array}{ccc}
\epsfxsize=7cm \epsffile{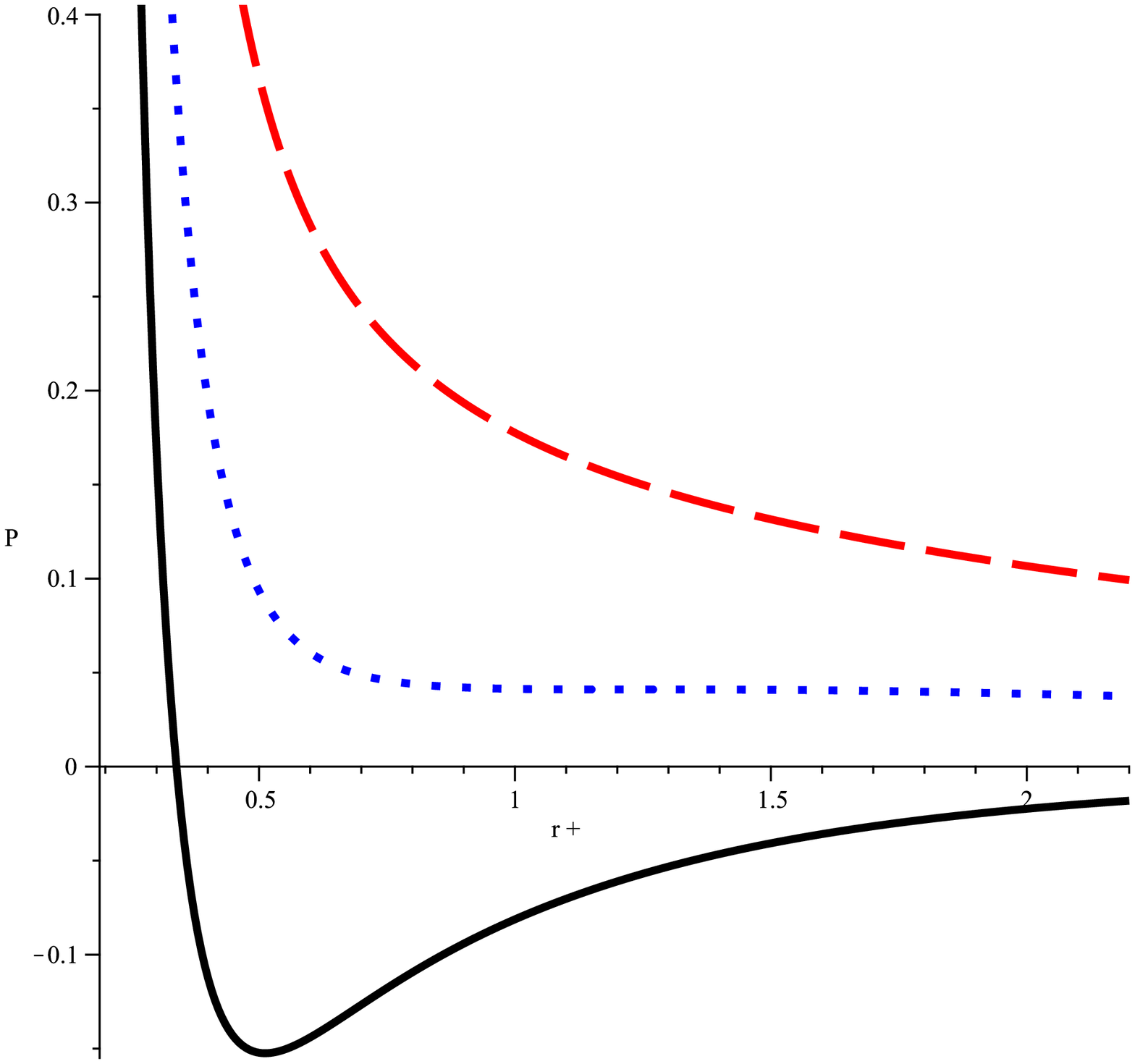} & \epsfxsize=7cm
\epsffile{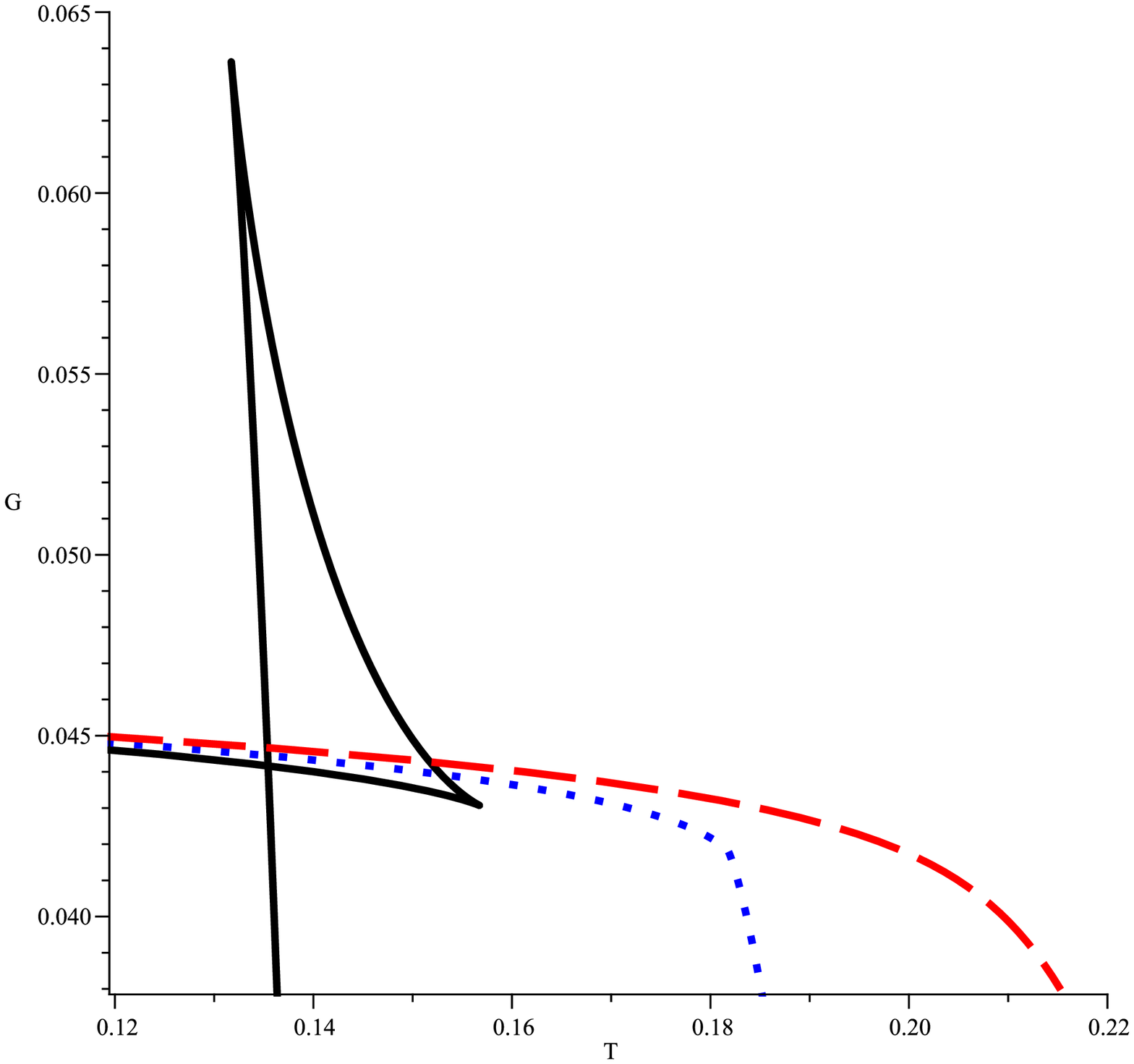} &
\end{array}
$%
\caption{$BD-BI: P-r_{+}$ (left), $G-T$ (right) diagrams for $b=1$, $n=4$, $%
q=1$ , $\protect\beta=0.5$ and $\protect\omega =100$.\newline
$P-r_{+}$ diagram, from bottom to up $T=0.1T_{c}$, $T=T_{c}$ and $T=2T_{c}$
respectively.\newline
$G-T$ diagram, from bottom to up $P=0.5P_{c}$, $P=P_{c}$ and $P=1.5P_{c}$,
respectively.}
\label{Figw1n4}
\end{figure}
\begin{figure}[tbp]
$%
\begin{array}{ccc}
\epsfxsize=7cm \epsffile{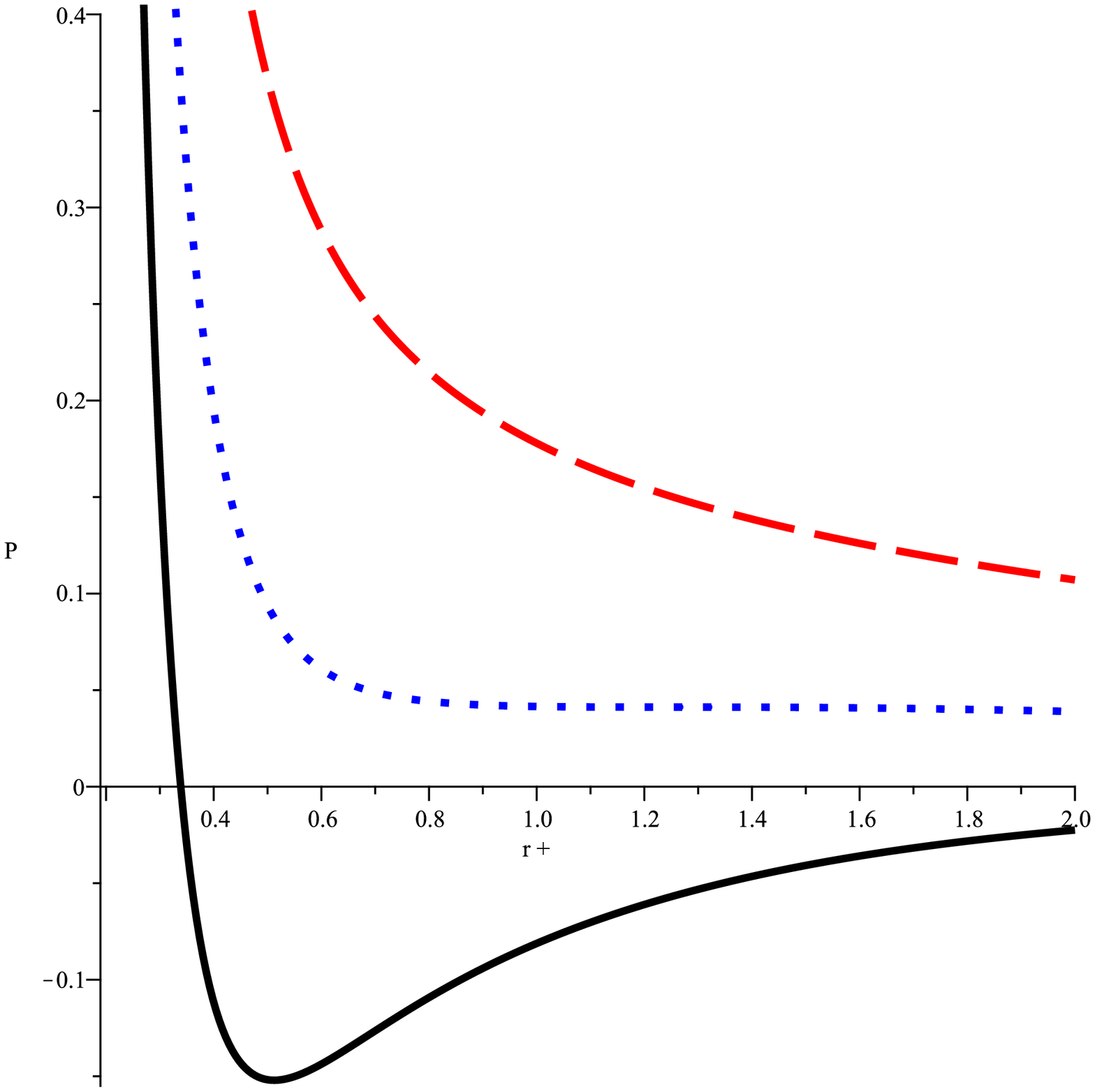} & \epsfxsize=7cm
\epsffile{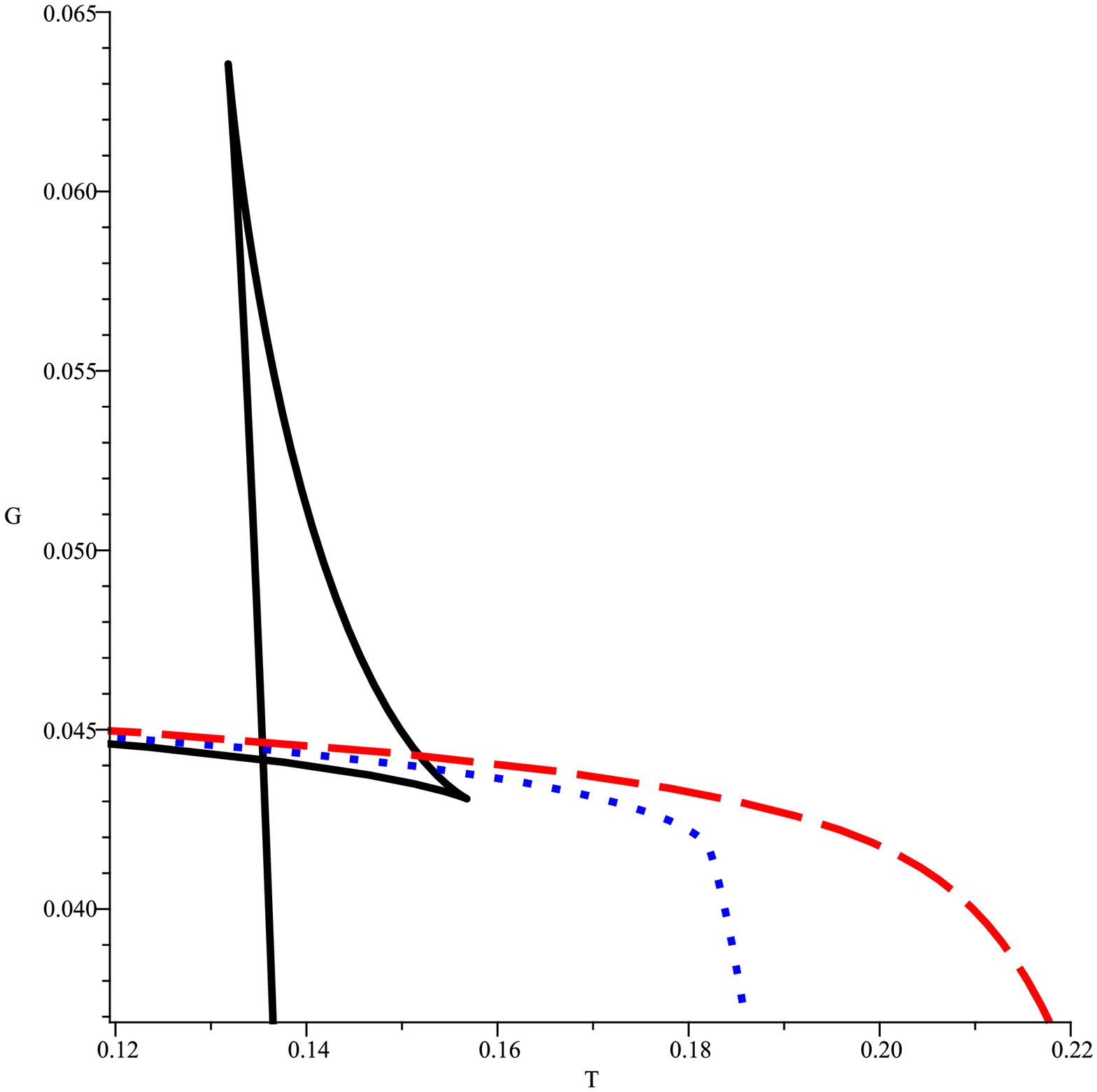} &
\end{array}
$%
\caption{$BI-dilaton: P-r_{+}$ (left), $G-T$ (right) diagrams for $b=1$, $%
n=4 $, $q=1$ , $\protect\beta=0.5$ and $\protect\omega
=100$.\newline $P-r_{+}$ diagram, from bottom to up $T=0.1T_{c}$,
$T=T_{c}$ and $T=2T_{c}$ respectively.\newline $G-T$ diagram, from
bottom to up $P=0.5P_{c}$, $P=P_{c}$ and $P=1.5P_{c}$,
respectively.} \label{Figw5n4}
\end{figure}

\begin{figure}[tbp]
$%
\begin{array}{ccc}
\epsfxsize=7cm \epsffile{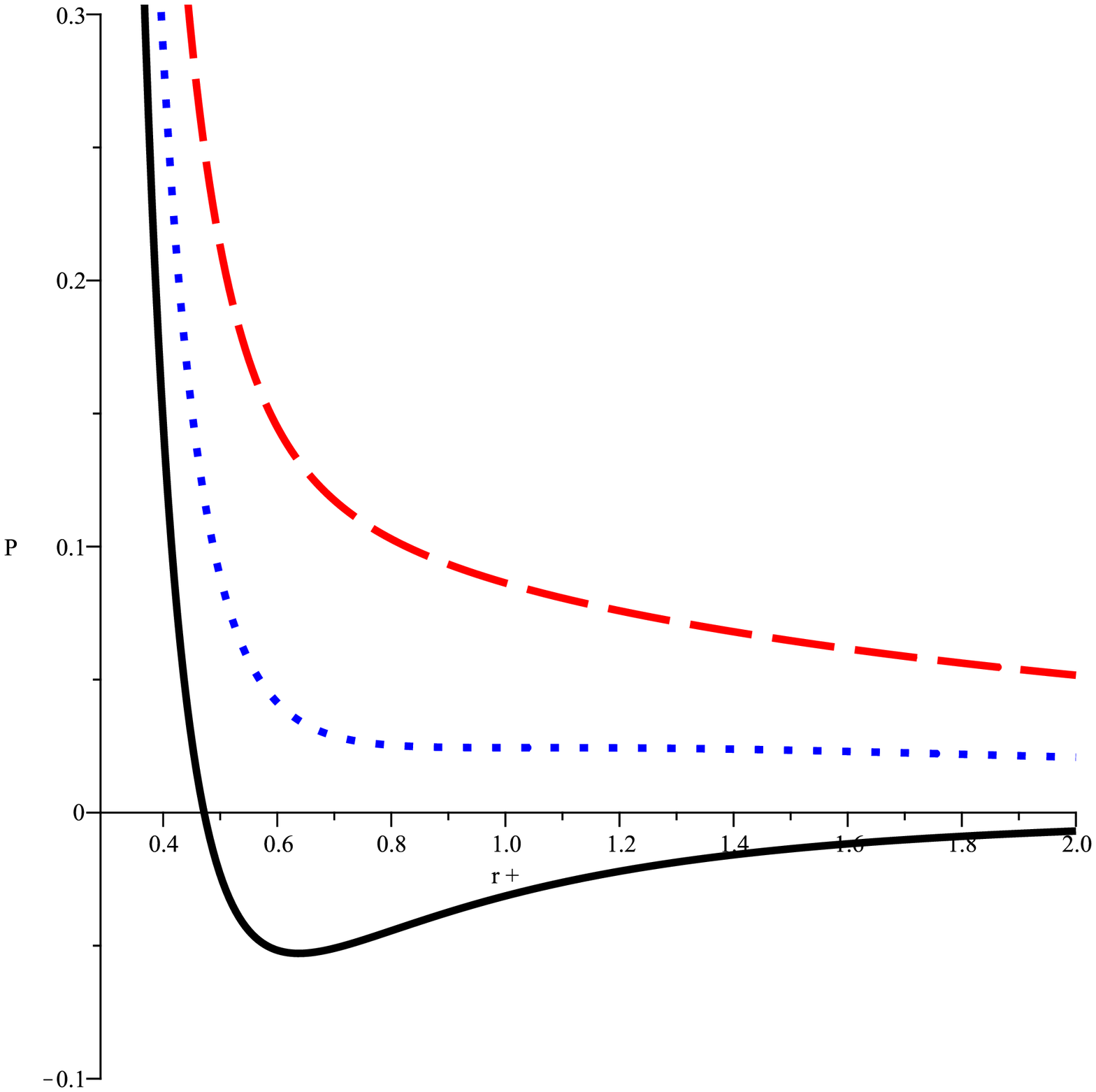} & \epsfxsize=7cm
\epsffile{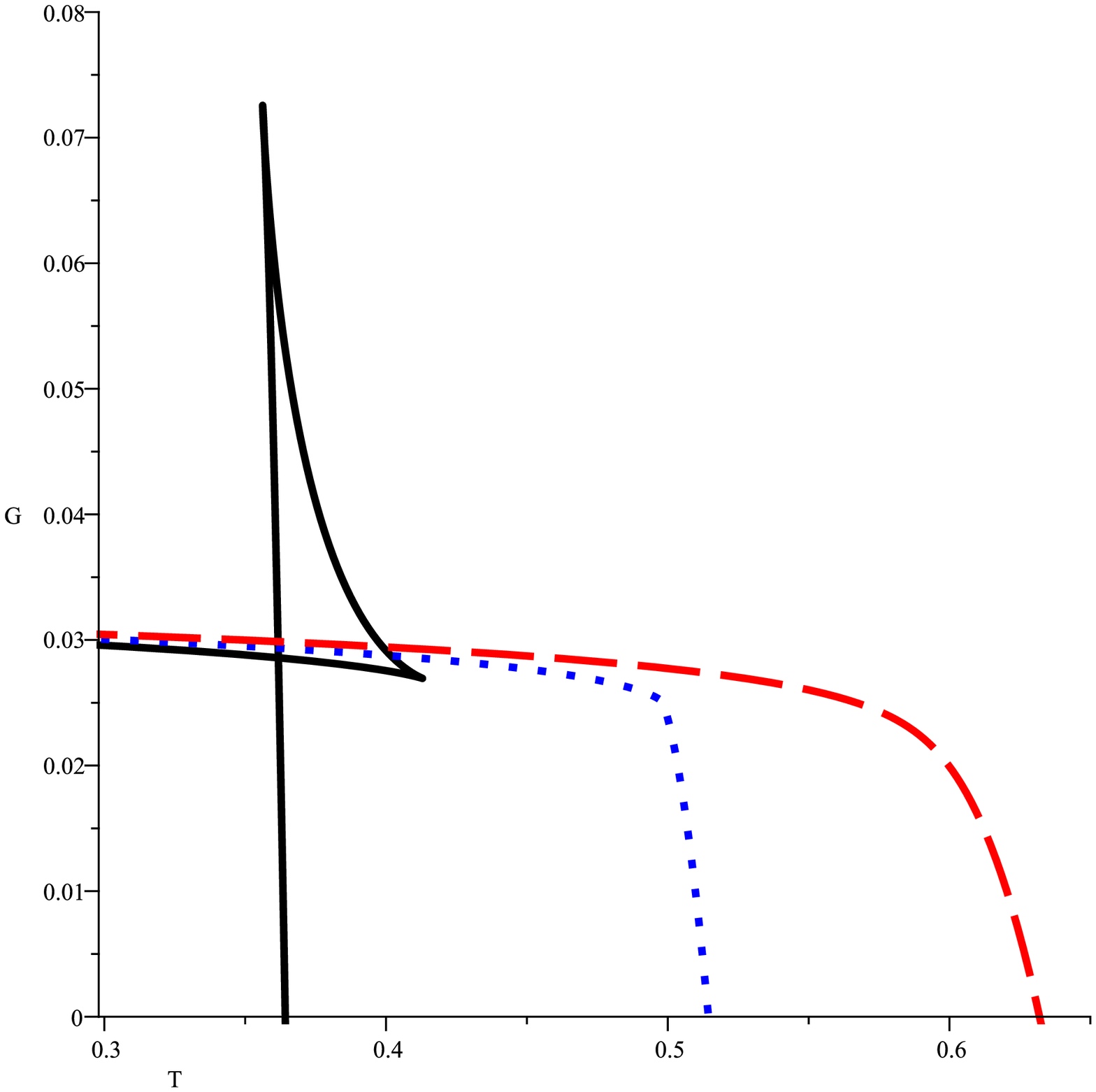} &
\end{array}
$%
\caption{$BD-BI: P-r_{+}$ (left), $G-T$ (right) diagrams for $b=1$, $n=6 $, $%
q=1$ , $\protect\beta=0.5$ and $\protect\omega =100$.\newline
$P-r_{+}$ diagram, from bottom to up $T=0.1T_{c}$ and $T=T_{c}$, $T=2T_{c}$
respectively.\newline
$G-T$ diagram, from bottom to up $P=0.5P_{c}$, $P=P_{c}$ and $P=1.5P_{c}$,
respectively.}
\label{Figw3n4}
\end{figure}

\begin{figure}[tbp]
$%
\begin{array}{ccc}
\epsfxsize=7cm \epsffile{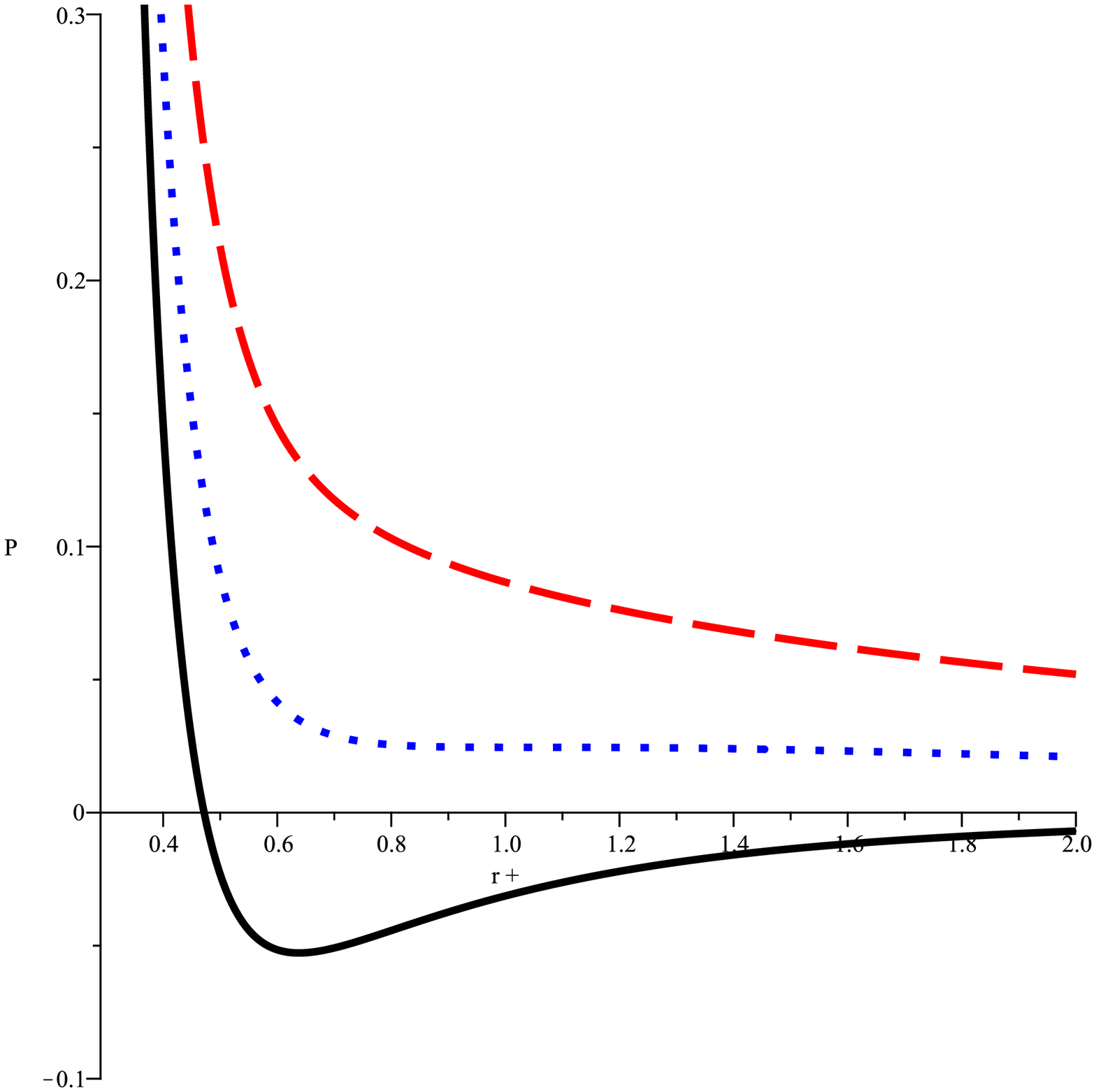} & \epsfxsize=7cm
\epsffile{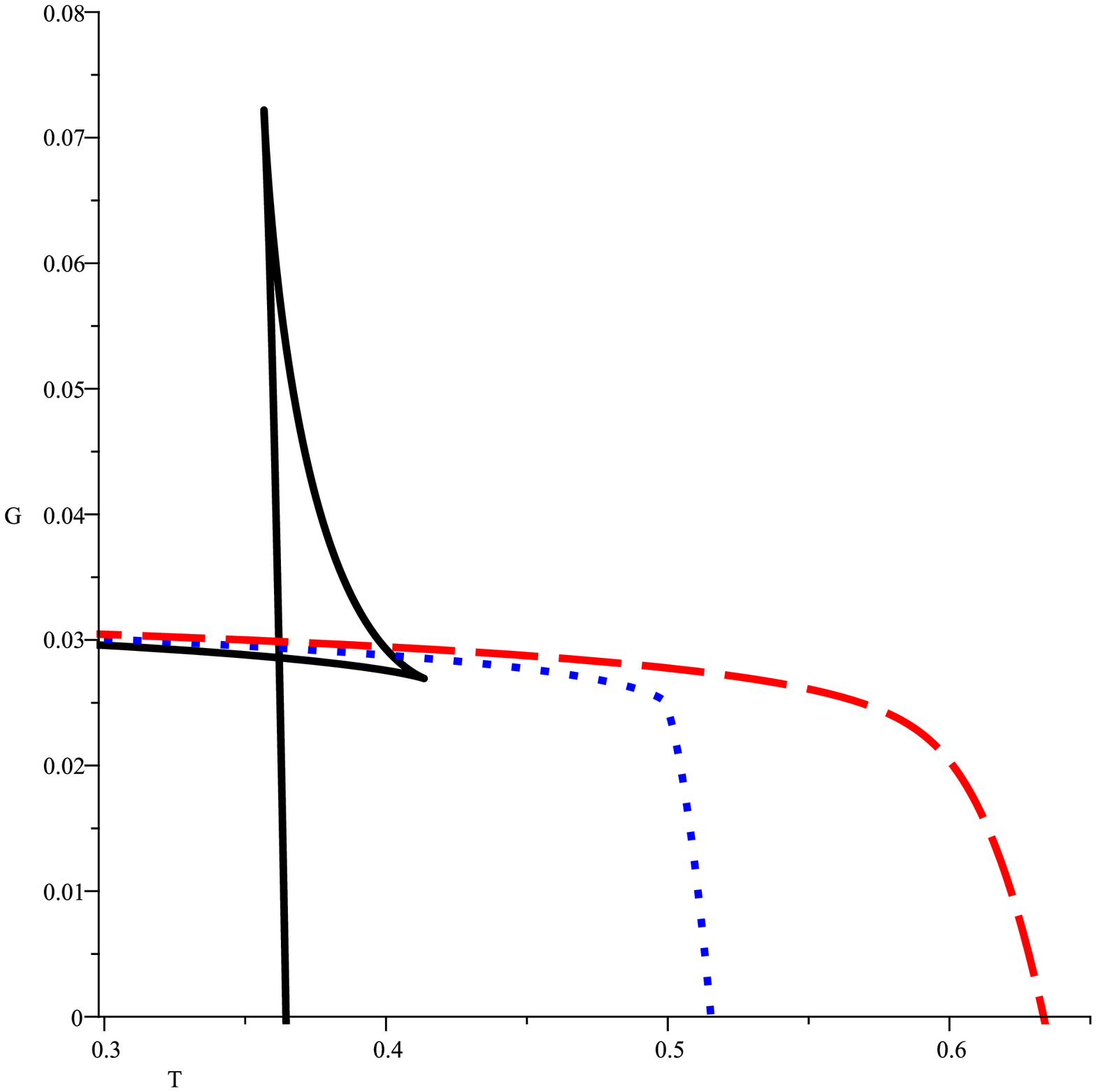} &
\end{array}
$%
\caption{$BI-dilaton: P-r_{+}$ (left), $G-T$ (right) diagrams for $b=1$, $%
n=6 $, $q=1$ , $\protect\beta=0.5$ and $\protect\omega
=100$.\newline $P-r_{+}$ diagram, from bottom to up $T=0.1T_{c}$,
$T=T_{c}$ and $T=2T_{c}$ respectively.\newline $G-T$ diagram, from
bottom to up $P=0.5P_{c}$, $P=P_{c}$ and $P=1.5P_{c}$,
respectively.} \label{Figw5n6}
\end{figure}

\begin{figure}[tbp]
$%
\begin{array}{ccc}
\epsfxsize=7cm \epsffile{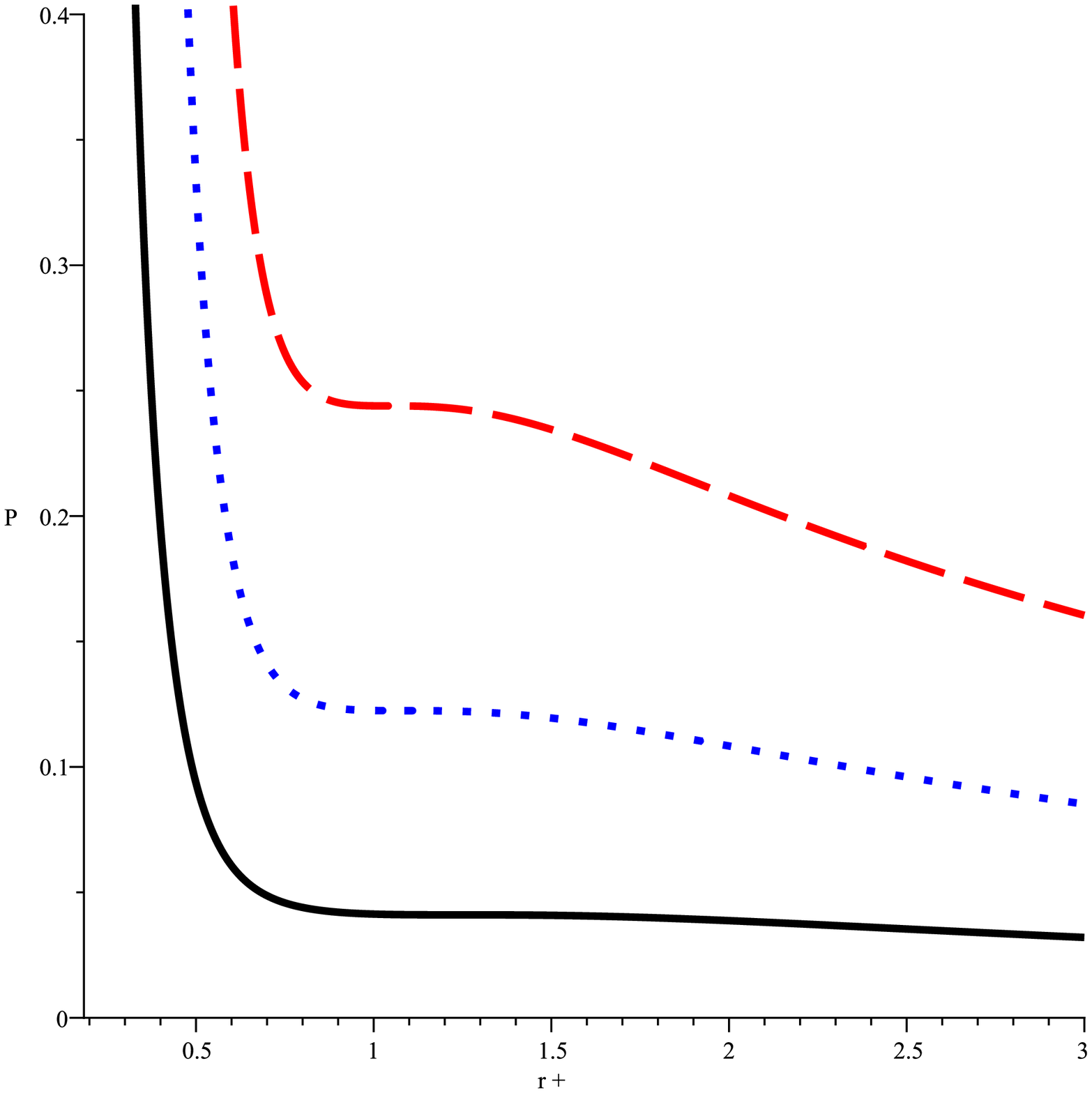} & \epsfxsize=7cm
\epsffile{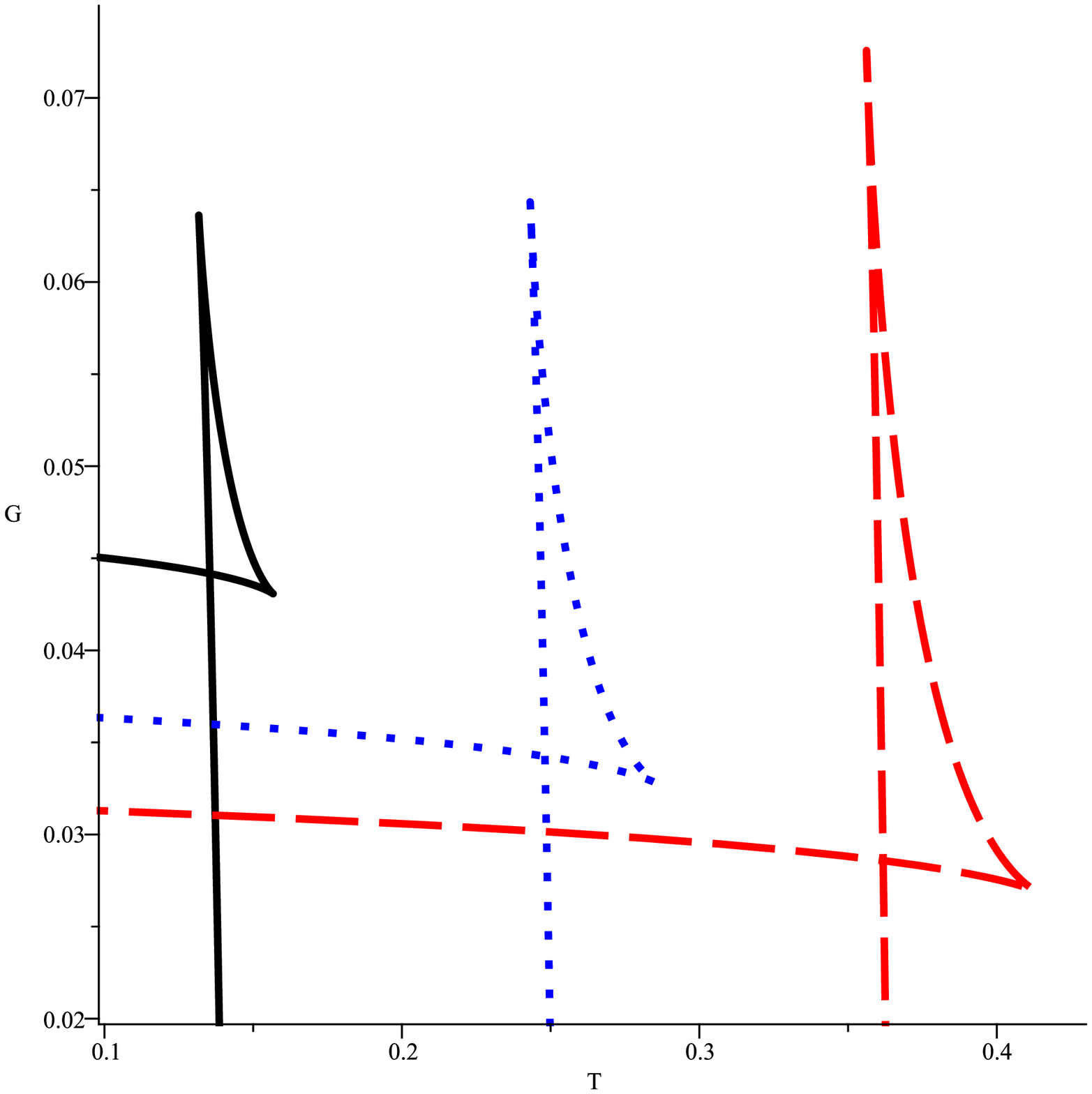} &
\end{array}
$%
\caption{$BD-BI: P-r_{+}$ (left), $G-T$ (right) diagrams for $b=1$, $\protect%
\omega =100$, $\protect\beta =0.5$, $q=1$.\newline $P-r_{+}$
diagram, for $T=T_{c}$, $n=4$ (solid line), $n=5$ (dotted line)
and $n=6$ (dashed line).\newline $G-T$ diagram, for $P=0.5P_{c}$,
$n=4$ (solid line), $n=5$ (dotted line) and $n=6$ (dashed line).}
\label{FignBD}
\end{figure}

\begin{figure}[tbp]
$%
\begin{array}{ccc}
\epsfxsize=7cm \epsffile{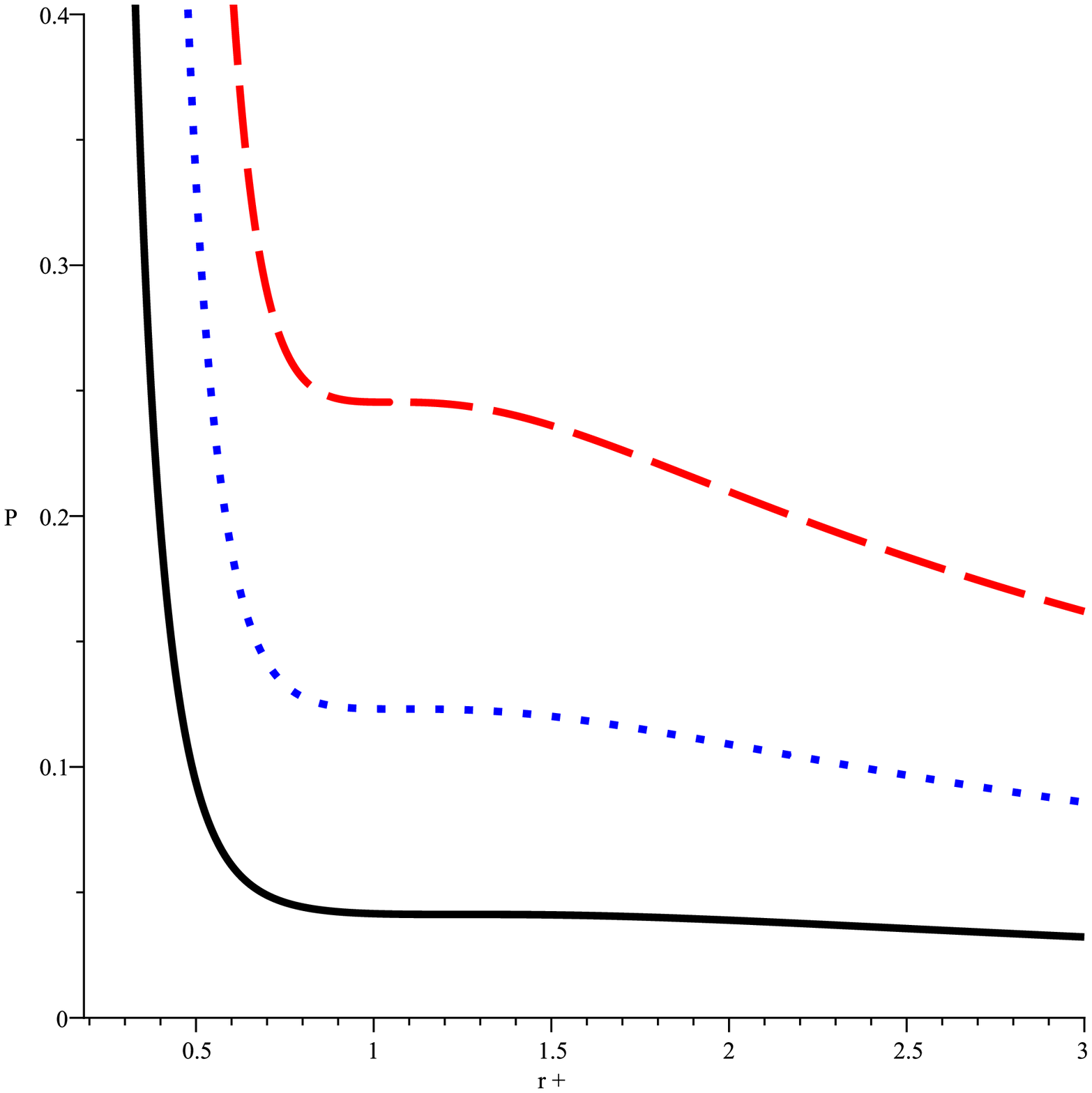} & \epsfxsize=7cm
\epsffile{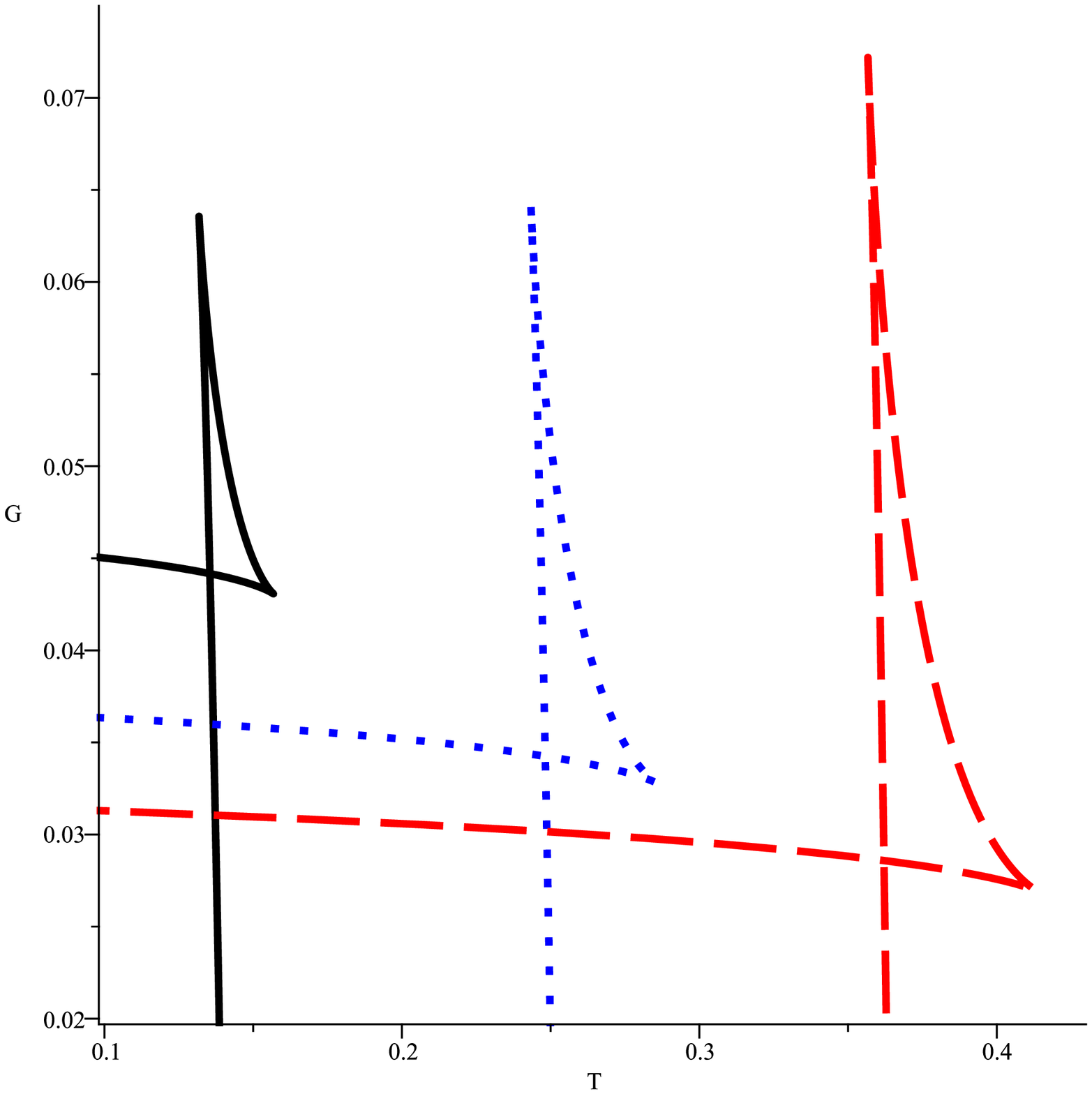} &
\end{array}
$%
\caption{$BI-dilaton: P-r_{+}$ (left), $G-T$ (right) diagrams for $b=1$, $%
\protect\omega =100$,$\protect\beta =0.5$, $q=1$.\newline
$P-r_{+}$ diagram, for $T=T_{c}$, $n=4$ (solid line), $n=5$
(dotted line) and $n=6$ (dashed line).\newline $G-T$ diagram, for
$P=0.5P_{c}$, $n=4$ (solid line), $n=5$ (dotted line) and $n=6$
(dashed line).} \label{Figndilaton}
\end{figure}

\begin{figure}[tbp]
$%
\begin{array}{ccc}
\epsfxsize=7cm \epsffile{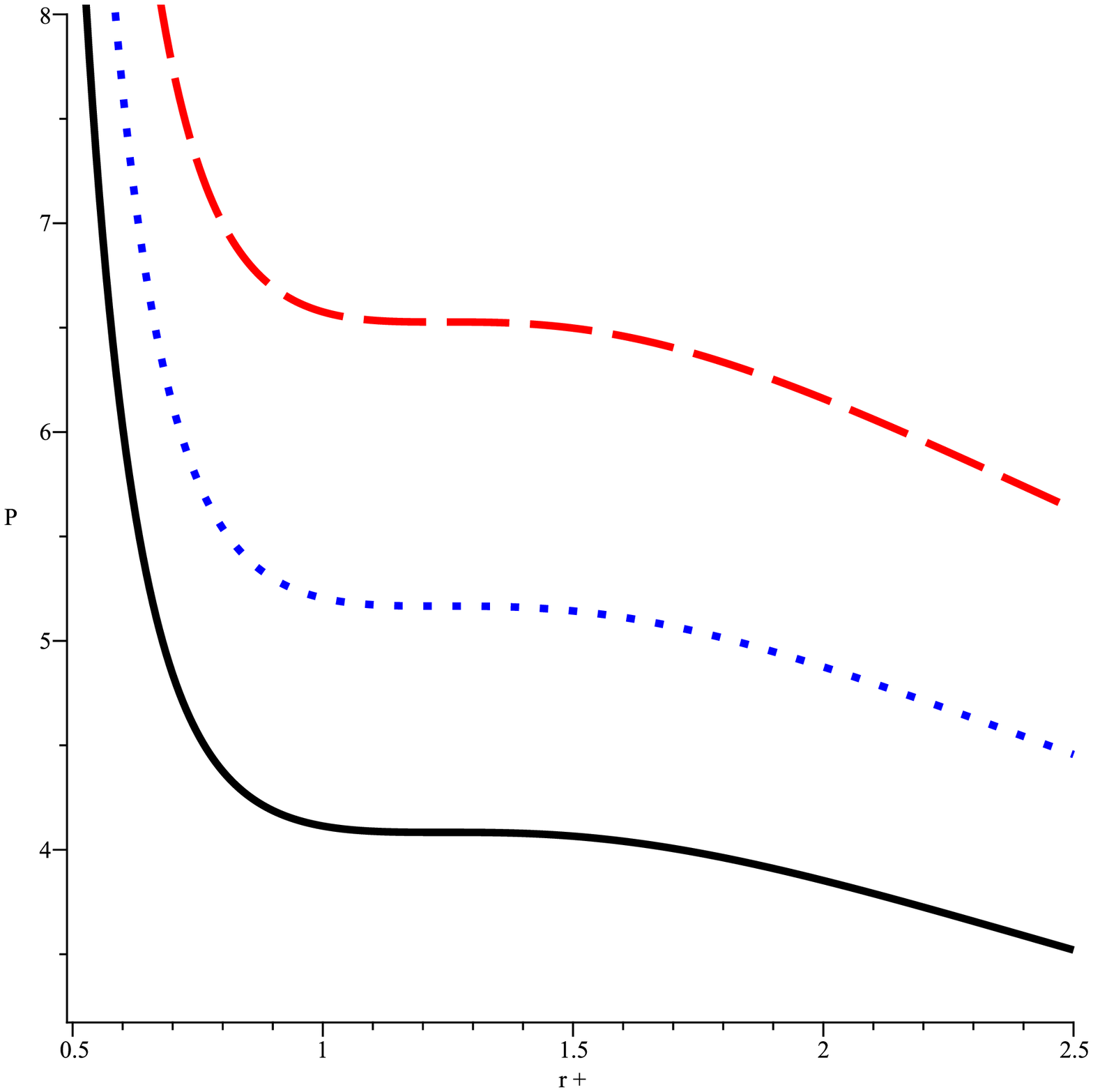} & \epsfxsize=7cm
\epsffile{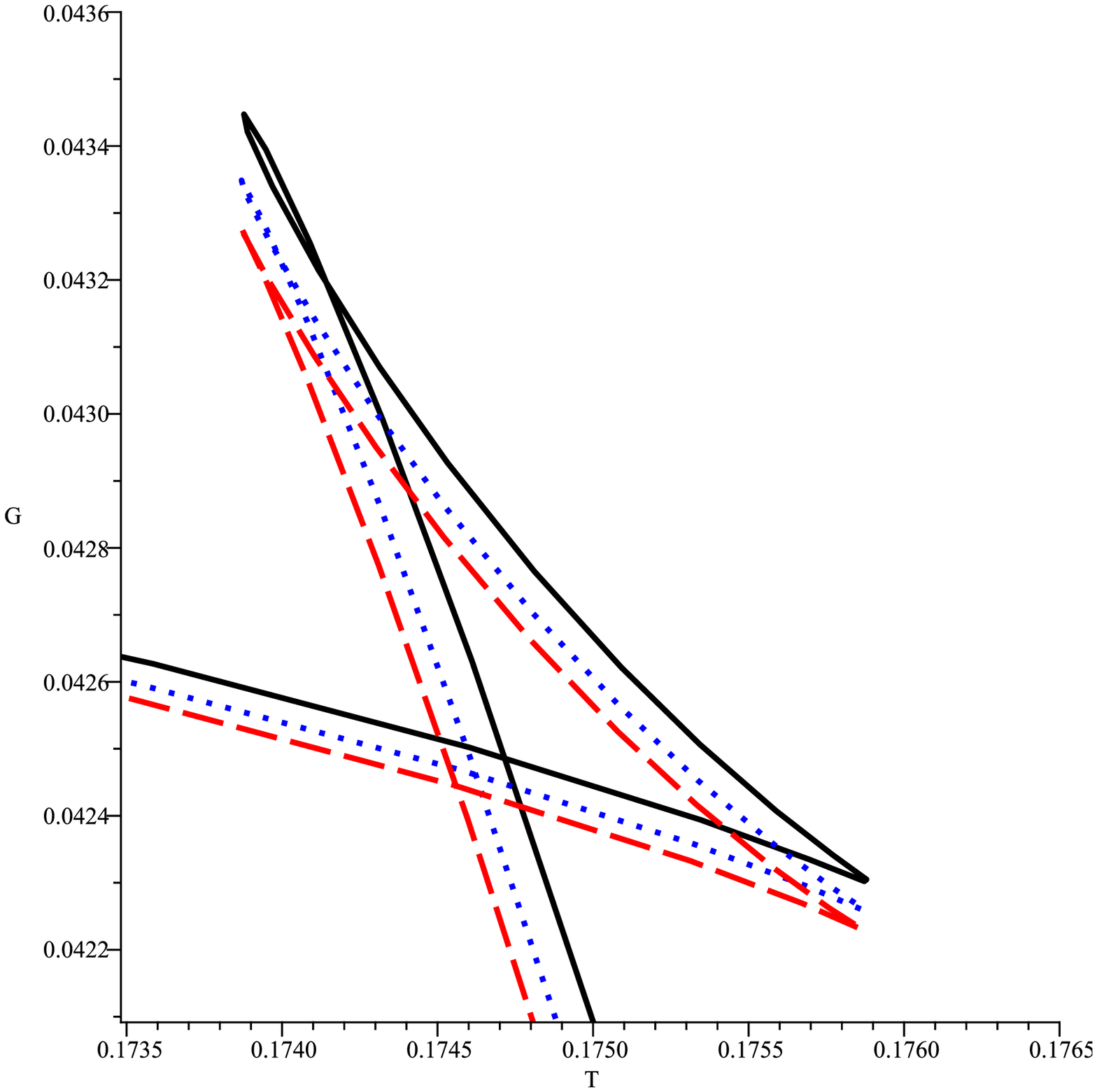} &
\end{array}
$%
\caption{$BD-BI: P-r_{+}$ (left), $G-T$ (right) diagrams for $b=1$, $n=4 $, $%
\protect\beta=0.5$ and $q=1$.\newline
$P-r_{+}$ diagram, for $T=T_{c}$, $\protect\omega =50$ (solid line), $%
\protect\omega =100$ (dotted line) and $\protect\omega =300$ (dashed line) .%
\newline
$G-T$ diagram, for $P=0.9P_{c}$, $\protect\omega =50$ (solid line), $\protect%
\omega =100$ (dotted line) and $\protect\omega =300$ (dashed line)
.} \label{Fig2n4BD}
\end{figure}

\begin{figure}[tbp]
$%
\begin{array}{ccc}
\epsfxsize=7cm \epsffile{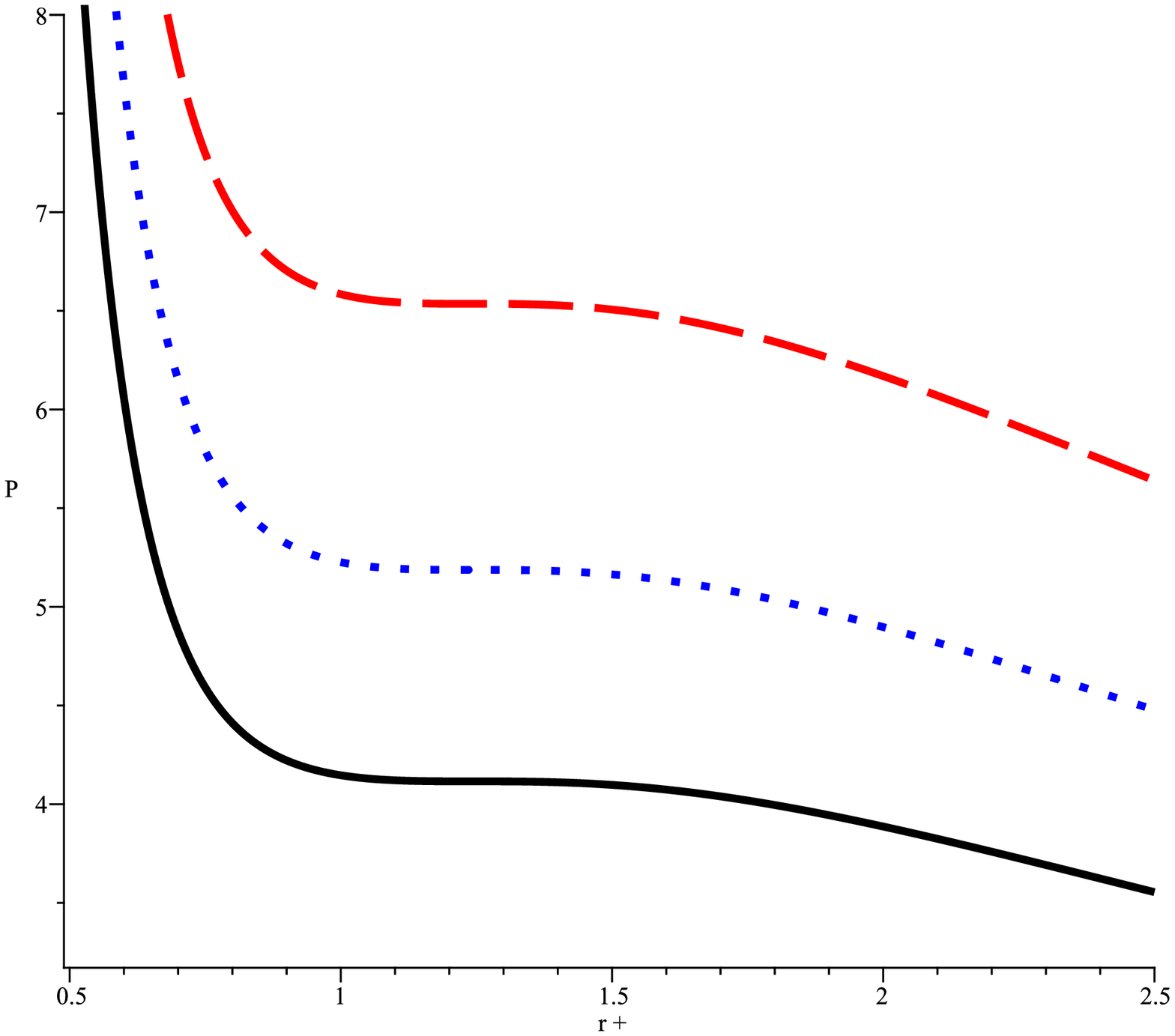} & \epsfxsize=7cm
\epsffile{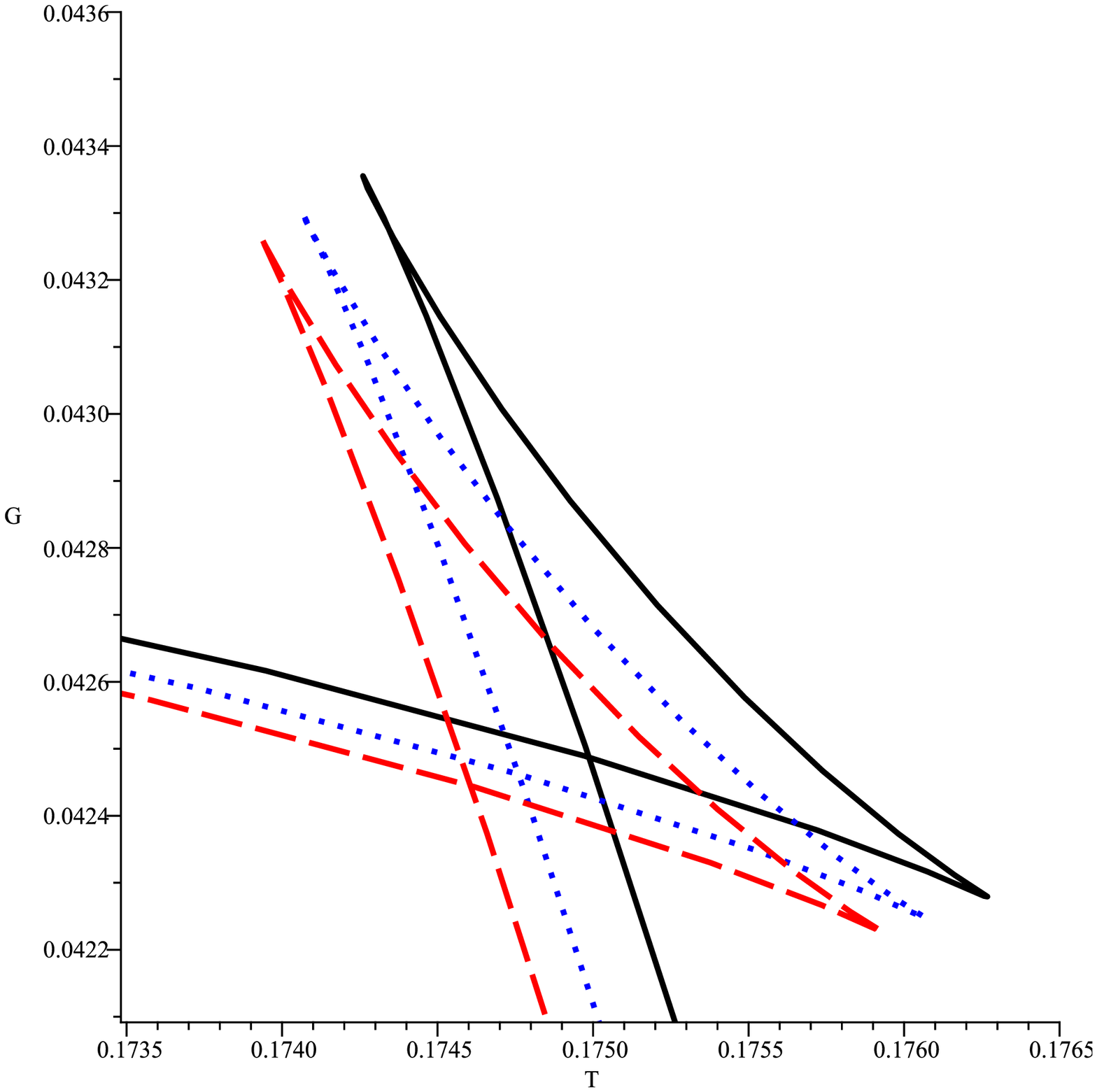} &
\end{array}
$%
\caption{$BI-dilaton: P-r_{+}$ (left), $G-T$ (right) diagrams for $b=1$, $%
n=4 $, $\protect\beta=0.5$ and $q=1$.\newline
$P-r_{+}$ diagram, for $T=T_{c}$, $\protect\omega =50$ (solid line), $%
\protect\omega =100$ (dotted line) and $\protect\omega =300$ (dashed line) .%
\newline
$G-T$ diagram, for $P=0.9P_{c}$, $\protect\omega =50$ (solid line), $\protect%
\omega =100$ (dotted line) and $\protect\omega =300$ (dashed line)
.} \label{Fig2n4d}
\end{figure}

\begin{figure}[tbp]
$%
\begin{array}{ccc}
\epsfxsize=7cm \epsffile{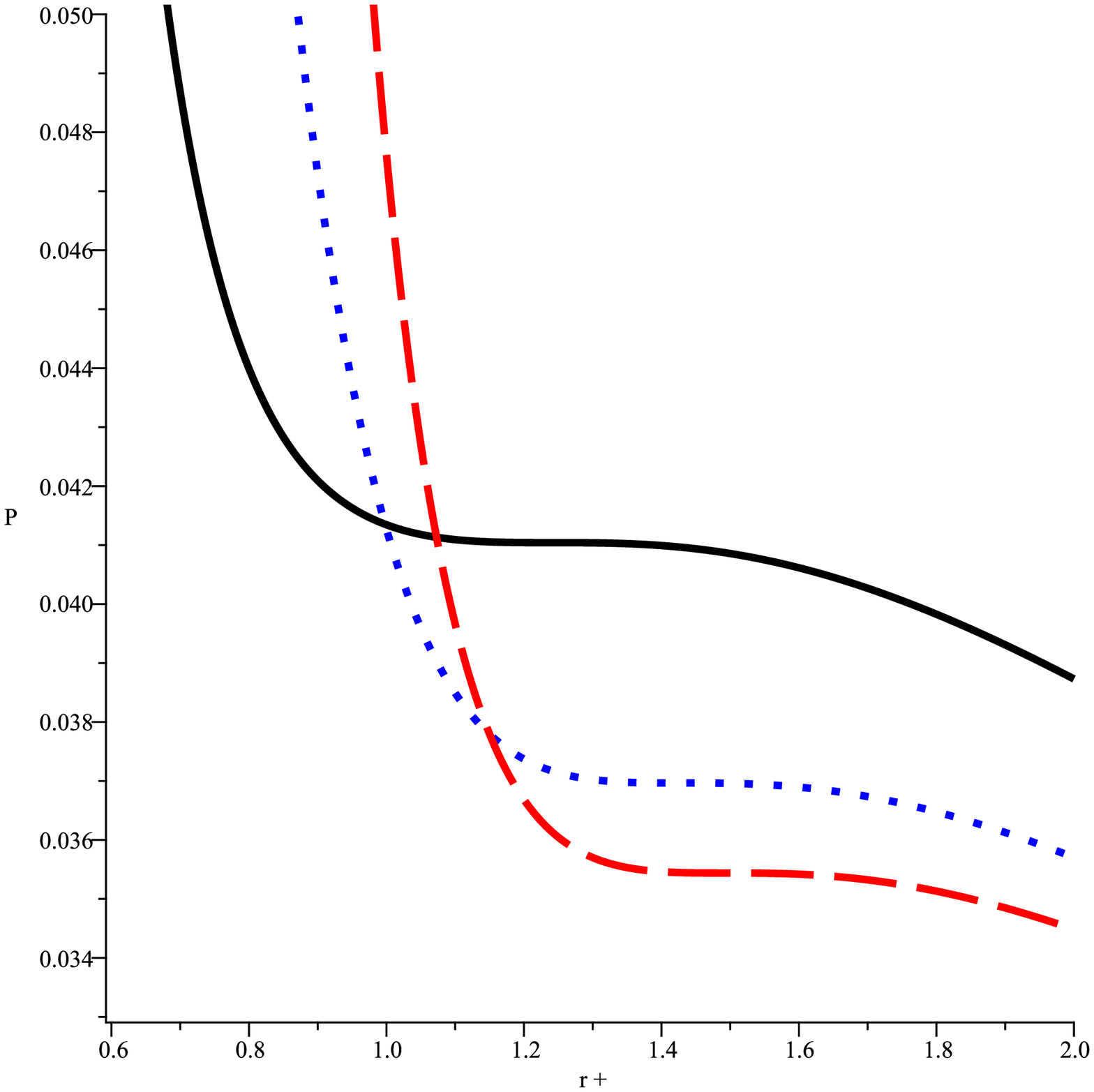} & \epsfxsize=7cm
\epsffile{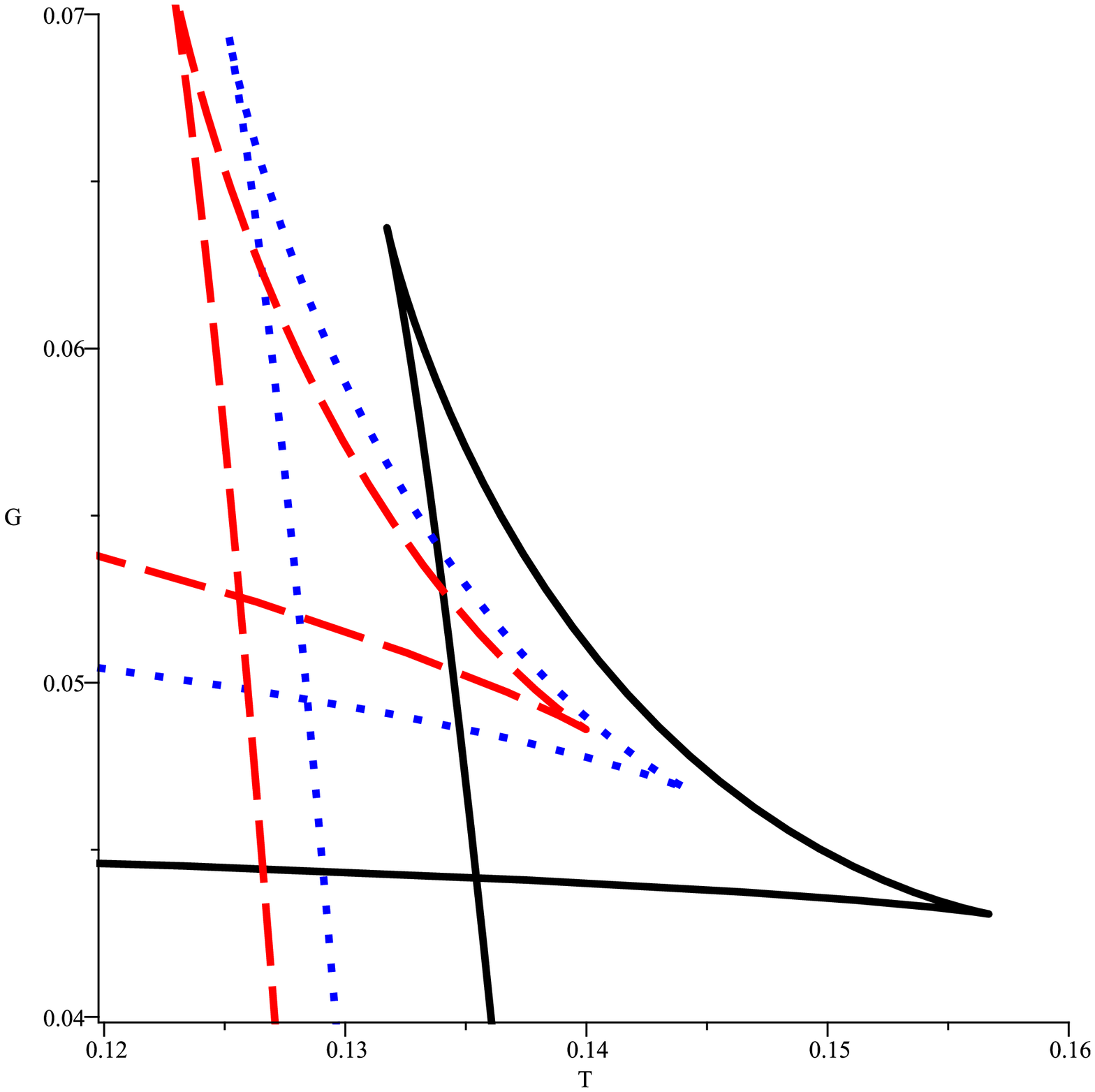} &
\end{array}
$%
\caption{$BD-BI: P-r_{+}$ (left), $G-T$ (right) diagrams for $b=1$, $n=4$, $%
\protect\omega=100$, $q=1$.\newline
$P-r_{+}$ diagram, for $T=T_{c}$, $\protect\beta=0.5$ (solid line), $\protect%
\beta=0.8$ (dotted line) and $\protect\beta=5$ (dashed line) .\newline
$G-T$ diagram, for $P=0.5P_{c}$ and $\protect\beta=0.5$ (solid line), $%
\protect\beta=0.8$ (dotted line) and $\protect\beta=5$ (dashed
line).} \label{FigbetaBD}
\end{figure}

\begin{figure}[tbp]
$%
\begin{array}{ccc}
\epsfxsize=7cm \epsffile{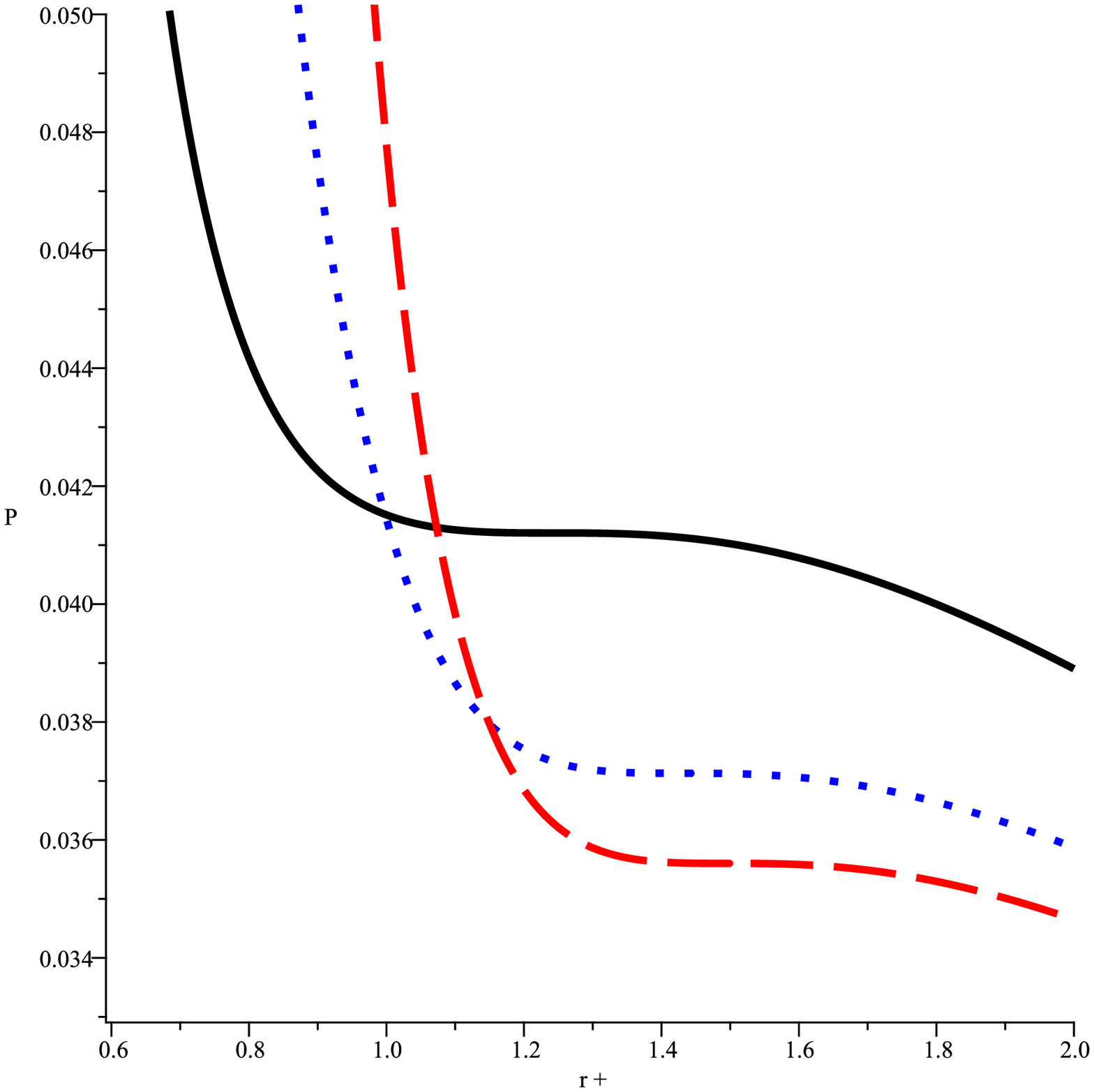} & \epsfxsize=7cm
\epsffile{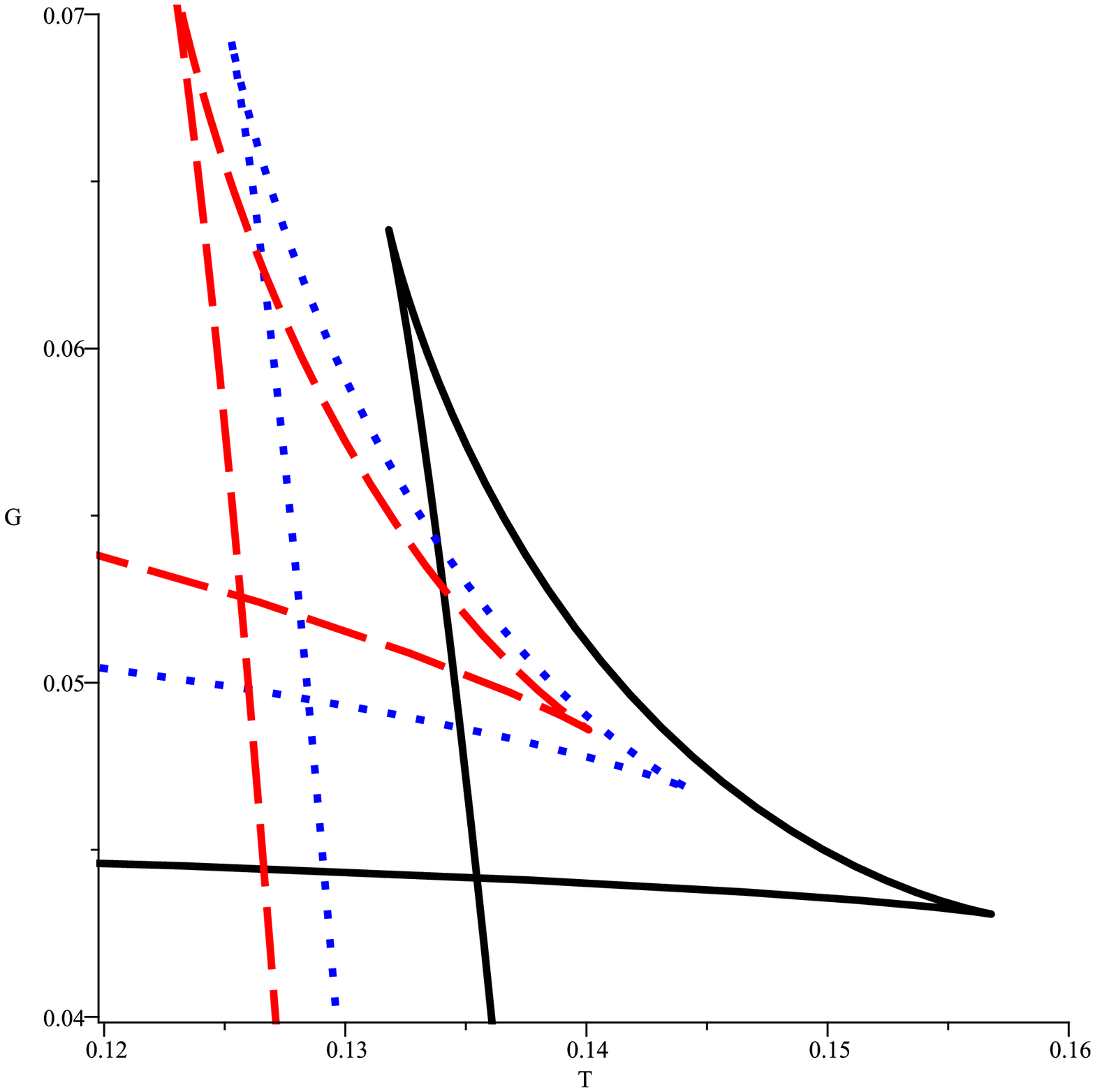} &
\end{array}
$%
\caption{$BI-dilaton: P-r_{+}$ (left), $G-T$ (right) diagrams for $b=1$, $%
n=4 $, $\protect\omega=100$, $q=1$.\newline
$P-r_{+}$ diagram, for $T=T_{c}$, $\protect\beta=0.5$ (solid line), $\protect%
\beta=0.8$ (dotted line) and $\protect\beta=5$ (dashed line) .\newline
$G-T$ diagram, for $P=0.5P_{c}$ and $\protect\beta=0.5$ (solid line), $%
\protect\beta=0.8$ (dotted line) and $\protect\beta=5$ (dashed
line).} \label{Figbetadilaton}
\end{figure}

\begin{figure}[tbp]
$%
\begin{array}{cc}
\epsfxsize=7cm \epsffile{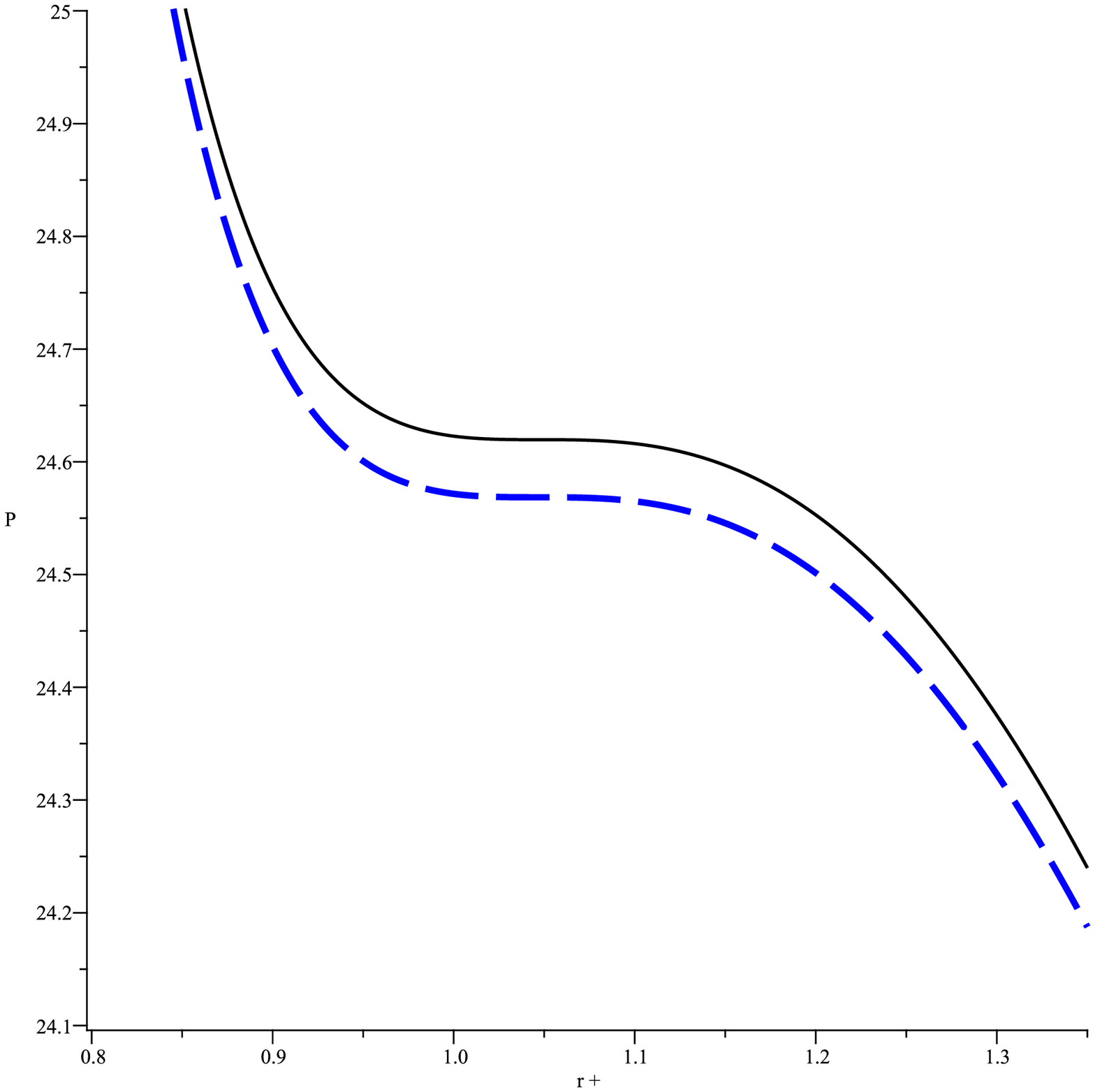} & \epsfxsize=7cm
\epsffile{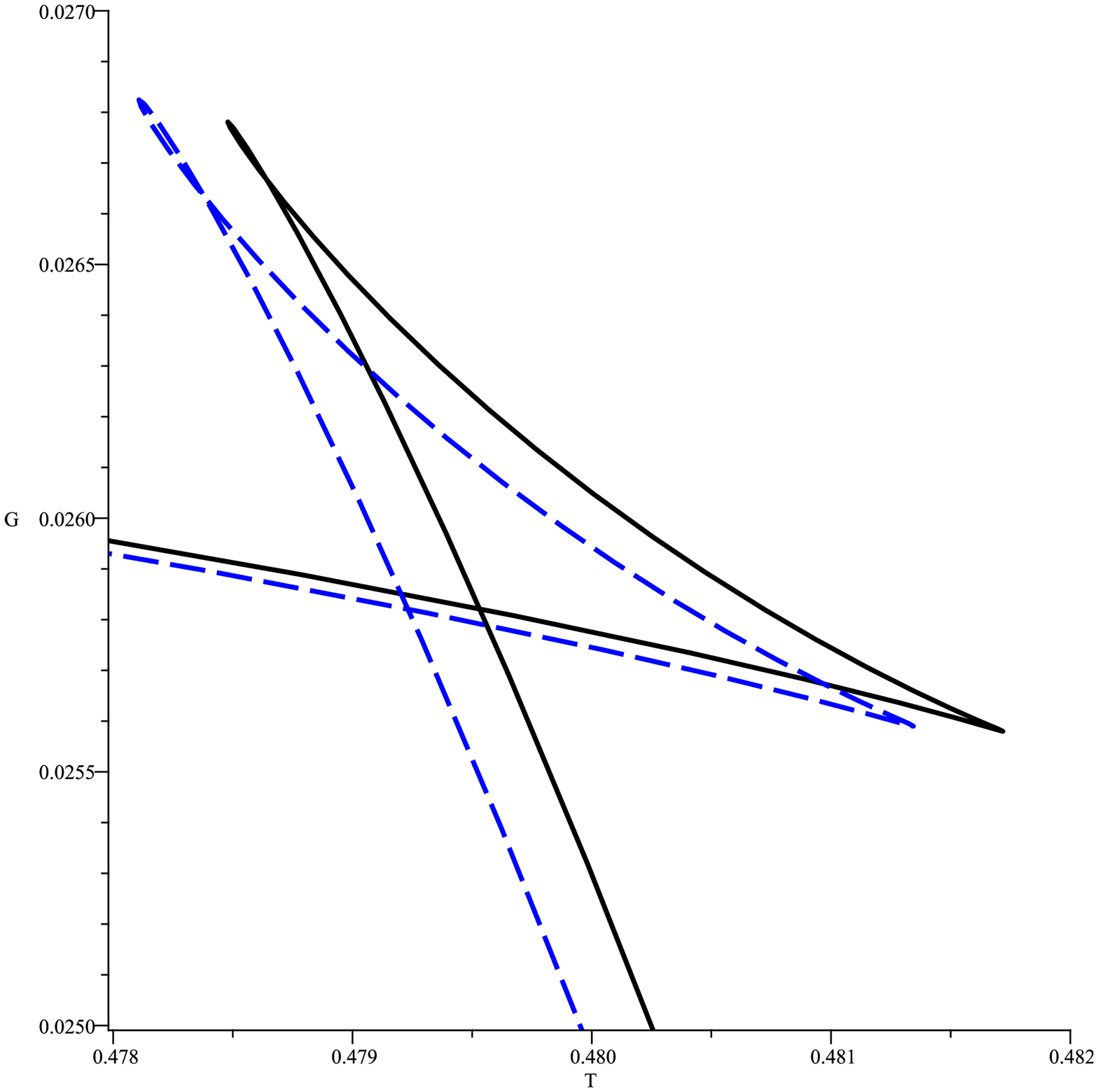}
\end{array}
$%
\caption{$P-r_{+}$ (left panel for $T=T_{c}$) and $G-T$ (right
panel for $P=0.92P_{c}$) diagrams for $b=1$, $n=6$,
$\protect\omega=300$, $q=1$ and $\beta=0.5$. BI-dilaton gravity
(solid line) and BD-BI gravity (dashed line).} \label{comparison}
\end{figure}


\begin{center}
\begin{tabular}{c}
\begin{tabular}{ccccc}
\hline\hline $\omega $ \;\;  & $r_{c}$ \;\;  & $T_{c}$ \;\;  & $P_{c}$\;\;  &$\frac{P_{c}r_{c}}{T_{c}}$ \\
\hline\hline \;\;\;\;$50.0000$ \;\;  & $1.240340$ \;\;& $0.181745$
\;\;& $0.040835$ \;\;  &$0.278683$ \\ \hline $100.0000$ \  &
$1.240487$ & $0.181762$ & $0.041042$ \  &$0.280102$ \\ \hline
$300.0000$ \  & $1.240588$ & $0.181775$ & $0.041183$ \  &$0.281068$ \\
\hline
\end{tabular}
\\
Table I: critical quantities of BD-BI for $q=1$, $\beta =0.5$ and
$n=4$.
\end{tabular}
\end{center}

\begin{center}
\begin{tabular}{c}
\begin{tabular}{ccccc}
\hline\hline $\omega $ \  & $r_{c}$ \  & $T_{c}$ \  & $P_{c}$ \;  &$\frac{P_{c}r_{c}}{T_{c}}$\\
\hline\hline $50.0000$ \;\;  & $1.242259$\;\; & $0.182185$\;\; &
$0.041158$ \;\;  &$0.280643$\\ \hline $100.0000$ \  & $1.241457$ &
$0.181986$ & $0.041206$ \;  &$0.281096$\\ \hline $300.0000$ \ &
$1.240914$ & $0.181850$ & $0.041238$ \;  &$0.281401$\\ \hline
\end{tabular}
\\
Table II: critical quantities of BI-dilaton for $q=1$, $\beta
=0.5$ and $n=4$.
\end{tabular}%
\end{center}

\begin{center}
\begin{tabular}{c}
\begin{tabular}{ccccc}
\hline\hline $n$ \  & $r_{c}$ \  & $T_{c}$ \  & $P_{c}$ \;  &$\frac{P_{c}r_{c}}{T_{c}}$\\
\hline\hline $4$\;\;  & $1.240487$ \;\; & $0.181762$ \;\; & $0.041042$ \;\;  &$0.280103$\\
\hline $5$ \  & $1.084670\ $ & $0.338047\ $ & $0.122431$ \;\;
&$0.392837$\\ \hline $6$ \  & $1.046170\ $ & $0.497403\ $ &
$0.243913$ \;\;  &$0.513014$\\ \hline
\end{tabular}
\\
Table III: critical quantities of BD-BI for $q=1$, $\beta =0.5$
and $\omega
=100$.%
\end{tabular}
\end{center}

\begin{center}
\begin{tabular}{c}
\begin{tabular}{ccccc}
\hline\hline $n$ \  & $r_{c}$ \  & $T_{c}$ \  & $P_{c}$ \;  &$\frac{P_{c}r_{c}}{T_{c}}$\\
\hline\hline $4$\;\;  & $1.241457$\;\; & $0.181986$\;\; & $0.041206$ \;\;  &$0.281096$\\
\hline $5$ \  & $1.085655\ $ & $0.338704\ $ & $0.123083$ \;\;
&$0.394521$\\ \hline $6$ \  & $1.047004\ $ & $0.498614\ $ &
$0.245419$ \;\;  &$0.527308$\\ \hline
\end{tabular}
\\
Table IV: critical quantities of BI-dilaton for $q=1$, $\beta =0.5$ and $%
\omega =100$.%
\end{tabular}%
\end{center}

\begin{center}
\begin{tabular}{c}
\begin{tabular}{ccccc}
\hline\hline $\beta $\  & $r_{c}$ \  & $T_{c}$ \  & $P_{c}$ \;  &$\frac{P_{c}r_{c}}{T_{c}}$\\
\hline\hline $0.5$\;\;  & $1.240488$\;\; & $0.181762$\;\; & $0.041042$ \;\;  &$0.280103$\\
\hline $0.8$ \  & $1.428906\ $ & $0.173651\ $ & $0.036966$ \;\;  &$0.304179$\\
\hline $5.0$ \  & $1.493876\ $ & $0.170385\ $ & $0.035439$ \;\;  &$0.310717$\\
\hline
\end{tabular}
\\
Table V: critical quantities of BD-BI for $q=1$, $n=4$ and $\omega =100$.%
\end{tabular}%
\end{center}

\begin{center}
\begin{tabular}{c}
\begin{tabular}{ccccc}
\hline\hline $\beta $ \  & $r_{c}$ \  & $T_{c}$ \  & $P_{c}$ \;  &$\frac{P_{c}r_{c}}{T_{c}}$\\
\hline\hline $0.5$\;\;  & $1.241457$\;\; & $0.181986$\;\; & $0.041206$ \;\;  &$0.281096$\\
\hline $0.8$ \  & $1.429402\ $ & $0.173883\ $ & $0.037131$ \;\;  &$0.305235$\\
\hline $5.0$ \  & $1.494287\ $ & $0.170617$ & $0.035603$ \;\;
&$0.311816$\\ \hline
\end{tabular}
\\
Table VI: critical quantities of BI-dilaton for $q=1$, $n=4$ and
$\omega =100
$.%
\end{tabular}%
\end{center}


\subsection{Discussion on the results of diagrams\label{plot}}

In Figs. $1-10$ we show the critical behavior of the system in
both BI-dilaton and BD-BI theories. In addition, we present six
tables to investigate the critical points, more clearly. As we
know, the phase transition occurs at the critical point, which
demonstrates the critical pressure, horizon radius and
temperature. Studying the $G-T$ diagrams of Einstein and Jordan
frames with related tables shows that by increasing the BD
coupling constant ($\omega$) (decreasing dilatonic coupling
constant $\alpha $ in Einstein frame), the critical values of
temperature and horizon radius increase (decrease) in Jordan
(Einstein) frame. Regarding $P-r_{+}$ diagrams and related tables,
we find that by increasing $\omega$, the critical pressure and
also the ratio $\frac{P_{c}r_{c}}{T_{c}}$ increase in both BD-BI
gravity its conformally related, BI-dilaton gravity. Comparing
right panels of Figs. \ref{Fig2n4BD} and \ref{Fig2n4d} with tables
I and II, we find an interesting result. According to the tables,
one finds increasing $\omega$ leads to increasing (decreasing)
$T_{c}$ in BD-BI (BI-dilaton) gravity. While regarding right
panels of Figs. \ref{Fig2n4BD} and \ref{Fig2n4d}, we find that for
$P<P_{c}$, the phase transition temperature is a decreasing
function in both frames.

We can also examine the effects of dimensionality on the critical
values and phase transition of the system in Figs. \ref{FignBD}
and \ref{Figndilaton} and related tables $III$ and $IV$. By
considering the $G-T$ and $P-r_{+}$ plots of Figs. \ref{FignBD}
and \ref{Figndilaton}, we can find that when $n$ increases, the
critical values of temperature and pressure, and the size of
swallow tail and the ratio $\frac{P_{c}r_{c}}{T_{c}}$ increase,
while the critical horizon radius decreases.

In addition, by studying tables $V$ and $VI$, we can find when the
nonlinearity parameter increases, the critical horizon radius and
the ratio $\frac{P_{c}r_{c}}{T_{c}}$ increase, but the critical
values of temperature and pressure decrease in both frames. One
can confirm this behavior in Figs. \ref{FigbetaBD} and
\ref{Figbetadilaton}.

To sum up, we can say that by increasing $\omega$, we have an
increment in critical value of pressure in both theories and a
reduction in the critical values of horizon radius and temperature
in dilaton gravity but in BD-BI gravity, they increase. On the
other hand, we can see that increasing the dimension leads to a
reduction in horizon radius and an increment in critical values of
pressure and temperature. In addition, by increasing $\beta$, we
have an increment in critical value of horizon radius and a
reduction in the critical values of pressure and temperature in
both frames. Furthermore regarding $G-T$ diagrams, it is clear
that although increasing $\beta$ leads to increasing $G$, the
Gibbs free energy decreases when we increase $n$ and $\omega$. It
is also notable that when we increase $\omega$, $n$ and $\beta$,
the ratio $\frac{P_{c}r_{c}}{T_{c}}$ increase too.  Furthermore,
we should note that although the related figures of both frames
are very similar, but the are not completely match to each other
(see Fig. \ref{comparison} for more clarifications). As a final
comment, we should note that for $\protect\omega \rightarrow
\infty$ ($\alpha \rightarrow 0$) and $\beta \rightarrow \infty$,
the solutions of BD-BI reduce to Reissner--Nordstr\"{o}m black
hole solutions. As a result, in order to have sensible effects for
BD and BI parameters, we should regard small values.

\section{Conclusions}

In this paper, the main goal was studying the properties of BD-BI
black hole solutions. At first, we have given a brief discussion
regarding to the old Lagrangian of BI-dilaton gravity. Since this
Lagrangian was not consistent with known BD-BI gravity, we had to
define a new Lagrangian. This new BI-dilaton Lagrangian emanates
from conformal transformation and it is a well-behaved Lagrangian
for $\beta \rightarrow \infty$. We have obtained new field
equations and calculated exact black hole solutions of field
equations in both Einstein and Jordan frames.

We have considered both BD-BI and BI-dilaton theories with
spherically symmetric horizon and studied their phase structure.
By considering cosmological constant proportional to
thermodynamical pressure and its conjugate variable as volume, we
have investigated the extended phase space and used the
interpretation of total mass of black hole as the enthalpy.

Studying calculated critical values through two different types of
phase diagrams (related to both frames) resulted in phase
transition taking place in the critical values. Studying $P-r_{+}$
and $G-T$ diagrams presented similar Van der Waals behavior near
the critical point. We have also shown that both scalar field and
nonlinearity parameter of electromagnetic field have considerable
effects on the critical quantities. We have examined the effects
of BD coupling constant, nonlinearity parameter and dimensionality
on the critical quantities. We have found that increasing BD
coupling constant leads to an increment in all critical quantities
in Jordan frame. In addition, for both frames, increasing
dimensionality leads to increasing critical values of temperature
and pressure, but decreasing the critical horizon radius. Also,
regarding the effects of $\beta$, we have found that its reduction
leads to reduction (increment) of critical pressure (critical
temperature and horizon radius). It is notable that although
changing $\omega$ does not have the same effects on some values of
critical quantities in both frames, variations of $n$ and $\beta$
have the same effects for both Einstein and Jordan frames. In
addition, we should note that increasing $\omega$, $n$ and $\beta$
leads to increasing the ratio $\frac{P_{c}r_{c}}{T_{c}}$,
regardless of $P_{c}$, $r_{c}$ and $T_{c}$ behaviors (for both
Einstein and Jordan frames).

Another interesting result of this paper is based on comparing the
consequences of both frames. Comparing the figures of BD-BI with
BI-dilaton branches, we have found that the total behavior of the
figures are very similar to each other. This is due to the fact
that the conformal factor is a regular smooth function at the
horizon and most of thermodynamic quantities are the same for both
frames. But as we have found in some figures (specially Fig.
\ref{comparison}) and related tables, the critical quantities are
not exactly the same in both frames.

Following the same adapted approach, the extended phase space and
$P-V$ criticality conditions of other models of nonlinear
electrodynamics are under investigation. In addition, the phase
transition of such solutions can be studied through geometrical
thermodynamics. This issue will be addressed in future work.

\begin{acknowledgments}
We would like to thank the anonymous referee for valuable
suggestions. We also acknowledge A. Poostforush and S. Panahiyan
for reading the manuscript. We thank the Research Council of
Shiraz University. This work has been supported financially by
Research Institute for Astronomy and Astrophysics of Maragha
(RIAAM), Iran.
\end{acknowledgments}

\end{document}